\documentclass[journal]{IEEEtran}  
 \IEEEoverridecommandlockouts                              

\usepackage{graphicx} 
\usepackage{enumerate}
\usepackage{subcaption}
\usepackage{cite}

\usepackage{bm}
\usepackage{mathtools}
\mathtoolsset{showonlyrefs}
\usepackage[ruled,vlined]{algorithm2e}

\usepackage{amsmath, amssymb, amsfonts, enumerate}
\usepackage{mathtools}

\DeclarePairedDelimiter\floor{\lfloor}{\rfloor}
\usepackage{cuted}

\def\reals{\mathbb R}

\def\ones{\mathbf{1}}

\def\Tcal{\mathcal T}

\newtheorem{theorem}{Theorem}
\newtheorem{lemma}{Lemma}
\newtheorem{proposition}{Proposition}
\newtheorem{corollary}{Corollary}

\newtheorem{remark}{Remark}
\newtheorem{definition}{Definition}
\newtheorem{assumption}{Assumption}
\usepackage[usenames,dvipsnames,table]{xcolor}

\usepackage{tikz, pgfplots}
\title{
Mobile Energy Storage in Power Network: Marginal Value and Optimal Operation
}

\author{Utkarsha~Agwan,%
~\IEEEmembership{Student Member,~IEEE,}
Junjie~Qin,%
~\IEEEmembership{Member,~IEEE,}
        Kameshwar~Poolla,%
        ~\IEEEmembership{Fellow,~IEEE,}
and~Pravin~Varaiya,%
~\IEEEmembership{Life~Fellow,~IEEE}%
\thanks{U. Agwan, K. Poolla and P. Varaiya are with the Department of Electrical Engineering and Computer Science, University of California at Berkeley
        {\tt\small \{uagwan, poolla, varaiya\}@berkeley.edu}}. 
\thanks{J. Qin is with the School of Electrical and Computer Engineering, Purdue University {\tt\small jq@purdue.edu}}%
\thanks{This work was supported in part by the C3.ai Digital Transformation Institute.}
}

\begin{document}

\IEEEoverridecommandlockouts
\IEEEpubid{\text{\begin{minipage}[t]{\textwidth}\ \\[10pt]
{A shorter version of this article is under review at IEEE Transactions on Smart Grids. An earlier version of this work was presented at the $60^{\text{th}}$ IEEE Conference on Decision and Control (CDC) in 2021. }
\end{minipage}}}

\maketitle


\begin{abstract}
This paper examines the marginal value of mobile energy storage, i.e., energy storage units that can be efficiently relocated to other locations in the power network. In particular, we formulate and analyze the joint problem for operating the power grid and a fleet of mobile storage units. We use two different storage models: \emph{rapid storage}, which disregards travel time and power constraints, and \emph{general storage}, which incorporates them. By explicitly connecting the marginal value of mobile storage to locational marginal prices (LMPs), we propose efficient algorithms that only use LMPs and transportation costs to optimize the relocation trajectories of the mobile storage units. Furthermore, we provide examples and conditions under which the marginal value of mobile storage is strictly higher, equal to, or strictly lower than the sum of marginal value of corresponding stationary storage units and wires. We also propose faster algorithms to approximate the optimal operation and relocation for our general mobile storage model, and illustrate our results with simulations of an passenger EV  and a heavy-duty electric truck in the PJM territory.
\end{abstract}

\def\Ds{\Delta^\mathrm{S}}
\def\Dm{\Delta^\mathrm{M}}
\def\eb{\mathbf{e}}

\section{Introduction}
With the increasing penetration of variable renewable energy resources, a fast growing amount of energy storage will be connected to the electric power network in the future. This is driven by profound technological and economic trends including the dramatic reduction in cost of battery systems and the advance of power electronics for programmable inverters. Meanwhile, mandates and goals for storage deployment have been proposed in many states including California and Massachusetts. As a result, new storage capacity is being connected to the grid at a record pace \cite{GTMstorage20}. On the other hand, transportation electrification and more generally the trend to ``electrify everything''\cite{8386924} are  expected to significantly increase the peak load for power systems. When the peak load reaches transmission or distribution capacity, the conventional wisdom is to install more wires or reinforce existing ones \cite{GACITUA2018346}. However, such wire-based solutions  have high capital costs, are usually time-consuming to implement, and can  face public opposition \cite{NRRINW}. Therefore, non-wire alternatives that can avoid or delay the need for capacity expansion are highly valued by system operators and utility companies. 

Mobile energy storage can simultaneously serve the role of energy storage and wires as it can help balance the supply and demand in both time and space. Mobile energy storage comes in many forms. Truck-mounted mobile energy storage units have been tested by Con Edison \cite{ConEd} for utility-scale applications. Electric vehicles and electric trucks together with bidirectional chargers can be used as mobile energy storage \cite{NissanLeaf}. Internet data centers whose load can be shifted across space (among data centers located at different geographical areas) and time \cite{10.1145/1993744.1993767} behave as a form of virtual mobile storage. Understanding the value of mobile storage, particularly with respect to the value of stationary storage and wires is an essential first step in assessing the potential of this technology for future power systems.

\subsection{Contribution and paper organization}
In this paper, we analyze the value of mobile storage from the point of view of a power system operator through a stylized formulation for joint operation of the power network and a fleet of mobile storage units. Given the relocation pattern of mobile storage, we show that the marginal value of mobile storage can be computed analytically using locational marginal prices. We first focus on \emph{rapid} mobile storage that can both charge/discharge instantly and travel instantly between nodes, and use the analytical expression for the marginal value of mobile storage to compare it with corresponding stationary storage and wires. Examples/conditions where the marginal value of mobile storage is strictly higher, lower, or equal to the sum of the marginal values of stationary storage and wires are presented, which offer insights on when mobile storage is preferred over stationary storage. We then propose efficient algorithms for the optimal storage relocation problem based on analytical expressions for the marginal value of \emph{rapid} mobile storage. We then extend our analysis to a \emph{general} mobile storage model which includes power constraints and travel time between nodes in the power network, and propose an algorithm to solve the optimal storage relocation problem. We also propose an approximate optimal storage relocation algorithm for the general mobile storage model to improve the solution time complexity. We illustrate our results with simulations of a passenger EV and an electric truck in a power network in the PJM territory. 

The paper is organized as follows: Section \ref{sec:model} describes our model for the power network, mobile storage and transport. Section \ref{sec:opt-dispatch} introduces the problem of multi-period economic dispatch with storage, and formulates the optimal relocation problem for mobile storage. Section \ref{sec:fast} studies the rapid mobile storage case, where storage can relocate and charge rapidly. Section \ref{sec:general} treats the general mobile storage case. Section \ref{sec:illustrations} illustrates our results with numerical examples, and Section \ref{sec:conclusion} concludes the paper. Due to space constraints, all proofs can be found in the appendix.

Preliminary results related to this paper appeared in the conference version \cite{agwan2021marginal}, where the value of mobile storage was quantified in terms of the LMPs, and algorithms were devised to optimally move and operate mobile storage. However, the algorithm proposed for the general mobile storage case had a time complexity which was exponential in the number of time steps in the decision horizon.
This paper significantly builds on work in \cite{agwan2021marginal} by devising an \emph{approximate} relocation algorithm which can be solved in polynomial time, whose solution converges to the true optimal solution, and by conducting case studies of passenger and heavy-duty electric vehicles providing mobile storage in a power network.


\subsection{Related Literature}

Several papers address the question of quantifying the marginal value of stationary storage.
Bose and Bitar \cite{bose2016marginal} develop an expression for locational marginal value of storage in a stochastic setting, i.e., when the system operator has to meet uncertain demand by dispatching generation and storage, and the marginal value is determined to be a function of the expected locational marginal prices (LMPs).
Qin et al. \cite{qin2018submodularity} derive an expression for the locational marginal value of stationary storage in terms of LMPs determined by the economic dispatch problem, and propose a discrete optimization framework for optimal siting from the system operator perspective.  Bitar et al. \cite{bitar2019marginal}  explore the economic value of storage colocated with a wind producer.
Each of these papers adds a level of complexity to the problem of valuing stationary storage, but considers an ideal storage model with no power constraints. Our results in Section \ref{sec:general}, when applied to stationary storage, generalize the state of the art stationary storage model by incorporating power constraints.

Compared to stationary storage, mobile storage can add extra value by relocating across the power network. 
Qin et al.\cite{QPV18MobileStorage1} propose using EV batteries to reduce demand charges, and justify the value proposition through a numerical case study. 
\cite{rossi2019interaction} models the coupling between transportation and power networks for on-demand transport services, and develop a joint optimization problem for the combined network. The economic value of truck mounted mobile storage is evaluated using numerical simulation in \cite{he2021utility}. 


\section{Modelling Grid, Storage, and Transportation}\label{sec:model}
\subsection{Notations}\label{sec:notations}
For any natural number $n$, let $[n]:= \{1, \dots, n\}$. We use $\mathbf{1}$ to denote all-one vectors of appropriate dimensions, $\eb_i \in \reals^n$ to denote the $i$-th elementary vector whose $i$-th element is one and all other elements are zeros. For any time dependent vector $\mathbf{x}(t)\in \reals^n$ with $t\in [T]$ and $T\in \mathbb{N}$, 
we use $\mathbf{x}_i\in \reals^T$ to denote $[x_i(1), \dots, x_i(T)]^\top$. 

\subsection{Power network}\label{sec:pn}
\def\pb{\mathbf{p}}
\def\elb{\bm{\ell}}
\def\gb{\mathbf{g}}
\def\fb{\mathbf{f}}
\def\fbb{\mathbf{\overline{f}}}
We consider the operation of a  power network with $n$ buses and $m$ lines over a finite horizon of length $T$. For each time slot $t\in [T]$, we denote the power generation and inelastic load over different buses in the power network by $\gb(t)\in \reals^n$ and $\elb(t)\in \reals^n$, respectively. The  generation cost function for generator at bus $i$ is denoted by $C_i(\cdot)$ and so the total generation cost in time period $t$ is 
\begin{equation}
	C(\gb(t)) = \sum_{i=1}^n C_i(g_i(t)), \quad t\in [T].
\end{equation}
For simplicity, we assume $C_i(\cdot)$ is strictly convex quadratic. This assumption can be replaced with other conditions that ensure differentiability of the optimal value function.

Denote the net power injection vector in time period $t$ by $\pb(t)\in \reals^n$. The net power injection vector needs to satisfy the linearized AC power flow constraints:
\begin{equation}
		\ones^\top \pb(t) =0, \quad 
		H \pb(t) \le \fbb, \quad t\in [T],
\end{equation}
where $\ones$ denotes the all-one vector,  $H\in \reals^{2m\times n}$ refers to the shift-factor matrix of the power network, and $\fbb\in \reals^{2m}$ models the thermal constraints of the lines (cf. \cite{qin2018submodularity} for a derivation of this version of the linearized AC power constraints). 

\def\ub{\mathbf{u}}
\def\ib{\mathbf{i}}
\subsection{Mobile energy storage}\label{sec:mes}
Consider a fleet of $K$ mobile energy storage units.  We denote the charging and discharging operation of storage unit $k$ in time period $t$ by $u_k(t)$, with the convention that $u_k(t)>0$ models charging and $u_k(t)<0$ models discharging.  We assume each storage unit starts empty at the beginning of the decision horizon, so the state of charge (SoC) of storage $k$ at the beginning of time period $t+1$ is $\sum_{\tau=1}^{t} u_k(\tau)$ for $k\in [K]$, $t\in [T]$.
The SoC of each storage unit must satisfy the energy capacity constraint. In particular, denote the energy capacity of storage $k$ by $\overline{s}_k$. Then we have 
\begin{equation}\label{eq:ec}
	0 \le \sum_{\tau=1}^{t} u_k(\tau) \le \overline{s}_k, \quad k\in [K], \quad t\in [T].
\end{equation}
Denote the vector of charging and discharging operation of storage $k$ by $\ub_k \in \reals^T$. Then constraint \eqref{eq:ec} can be written as 
\begin{equation}
	\mathbf{0} \le L\ub_k \le \overline{s}_k \ones,\quad k\in [K],
\end{equation}
where $L$ is a lower triangular matrix defined as $L_{ij} = 1$, if $i \ge j$, and $0$ otherwise.

In addition to the energy capacity constraint, storage units often have a power capacity rating which limits the amount of energy that can be charged or discharged for each unit of time. Denote the power rating of storage $k$ by $\overline{u}_k$. In general, when we vary the energy capacity of a storage units (e.g. by adding more battery packs), the power rating will also vary. Thus we write $\overline{u}_k= \overline{u}_k(\overline{s}_k)$ when we want to highlight this dependence, where $\overline{u}_k(\cdot)$ is continuously differentiable. \footnote{If $\overline{u}_k$ is determined by the power rating of the charger, $\overline{u}_k$ is a constant and $\frac{\partial \overline{u}_k}{\partial\bar{s}_k}$ is $0$.}

The amount of energy a storage unit can charge or discharge in a time period also depends on the time that is available for  storage  operation. 
Denote the length of each time period in our discrete time model by $\Delta$. Since it takes time for mobile storage units to move between different buses in the power network, the time that is available for storage $k$ to charge or discharge in time period $t$, denoted by $\Ds_k(t)$, satisfies $0\le \Ds_k(t) \le \Delta$. As a result, the power capacity leads to the following constraint
\begin{equation}\label{eq:pc}
	-  \Ds_k(t) \overline{u}_k \le u_k(t) \le \Ds_k(t) \overline{u}_k, \quad  k\in [K], \quad t\in [T]. 
\end{equation}

 Denote the bus at which storage $k$ is located at the beginning of time period $t$ by $i_k(t)$, and define matrix $E(t)\in \reals^{n\times K}$ as
 \begin{equation}\label{eq:Et}
 	E(t) = \begin{bmatrix}
 		\eb_{i_1(t)} & \dots & \eb_{i_K(t)}
 	\end{bmatrix}, \quad t\in [T],
 \end{equation}
 where $\eb_i$ is the $i$-th elementary vector with a one at the $i$-th element and zeros elsewhere. Note that the sequence $\{i_k(t)\}_{t\in [T]}$ characterizes the relocation process (i.e., the trajectory) of storage $k$, while $E(t)$ provides a snapshot of the locations of all  mobile storage units in the power network in time period $t$. We denote the collections of trajectories and snapshots by
 \begin{equation}
 	\mathfrak{I} =\{\mathfrak{I}_k\}_{t\in [T]} :=\{i_k(t)\}_{k\in [K], t\in [T]}, \quad \mathfrak{E}:= \{E(t)\}_{t\in [T]},
 \end{equation}
 respectively, where we use $\mathfrak{I}_k$ to denote the trajectory of mobile storage $k$.  Since \eqref{eq:Et} defines a one-to-one mapping between $\mathfrak{I}$ and $\mathfrak{E}$, we will use them interchangeably.

\subsection{Transportation model}\label{sec:tm}
The system operator can determine whether to relocate storage $k$ in each time period. If storage $k$ is relocated in time period $t$, a portion of time period $t$ is used to move the storage from one bus to another. Let $D\in \reals^{n\times n}$ be a matrix whose $(i,j)$-th element is the travel time from bus $i$ to bus $j$. We assume that the length of the time intervals $\Delta$ is selected so that $D_{i,j} \le \Delta$ for all $i,j$. 
 Denote the time needed for moving storage $k$  in time period $t$ by $\Dm_k(t)$. Since storage $k$ is moved from bus $i_k(t)$ to bus $i_k(t+1)$ in time period $t$, we have
 \begin{equation}\label{eq:Dm}
 	\Dm_k(t) = \eb_{i_k(t)}^\top D \eb_{i_k(t+1)}, \quad  k\in [K], \quad t\in [T]. 
 \end{equation}
 Thus the time left for storage to charge or discharge in time period $t$ is 
 \begin{equation}\label{eq:Ds}
 	\Ds_k(t) = \Delta - \Dm_k(t), \quad  k\in [K], \quad t\in [T]. 
 \end{equation}
We adopt the convention that if a storage $k$ is moved in time period $t$, it will first use $\Ds_k(t)$ time to charge/discharge at bus $i_k(t)$ and then use $\Dm_k(t)$ time to relocate to bus $i_k(t+1)$. 

Transporting energy storage between buses is costly. For example, truck mounted mobile storage consumes fuel to travel between buses. Denote the relocation cost per unit of travel time by $\kappa$. Then the total  cost for relocation is 
\begin{equation}\label{eq:fcost}
	J^{\mathrm{R}}(\mathfrak{I}):= \sum_{k=1}^K J_k^{\mathrm{R}}(\mathfrak{I}_k):= \kappa\sum_{k=1}^K \sum_{t=1}^T  \eb_{i_k(t)}^\top D \eb_{i_k(t+1)},
\end{equation} 
where $J_k^{\mathrm{R}}(\mathfrak{I}_k)$ is the relocation cost for the $k$-th mobile storage unit. 

\section{Optimal Dispatch and Relocation}\label{sec:opt-dispatch}
In a centralized optimization setting, the system operator controls both the power grid and the mobile storage fleet. Given the locations of the mobile storage units $\mathfrak{E}$, the operation of the grid and the charging/discharging of the storage units can be formulated as the following optimization problem:
\begin{subequations} \label{eq:ED} \noeqref{eq:ED-obj}
	\begin{align} 
		\min_{\mathbf{g}, \mathbf{u}} \quad & \sum_{t=1}^T C_t (\mathbf{g}(t)) \label{eq:ED-obj} \\
		\mbox{s.t.} \quad &  \mathbf{1}^\top(\mathbf{g}(t) - \mathbf{d}(t) - E(t)\mathbf{u}(t)) = 0,& \!\!\!\!t\in [T], \label{eq:ED-demand}\\
		&   H (\mathbf{g}(t) - \mathbf{d}(t) - E(t)\mathbf{u}(t)) \le \overline{\mathbf{f}}, & \!\!\!\!t\in [T], \label{eq:ED-line} \\
		&   \mathbf{0} \le L \mathbf{u}_k \le \overline{s}_k \mathbf{1}, & \!\!\!\!k \in [K], \label{eq:ED-energy}\\
		&  -  \overline{u}_k(\overline{s}_k)\bm{\Delta}^{\mathrm{S}}_k  \le \ub_k \le \overline{u}_k(\overline{s}_k)\bm{\Delta}^{\mathrm{S}}_k, & \!\!\!\! k\in [K]. \label{eq:ED-power}
	\end{align}
\end{subequations}
where $\{\bm{\Delta}^{\mathrm{S}}_k\}_{k\in [K]}$ is calculated via \eqref{eq:Dm} and \eqref{eq:Ds}, and the locations of mobile storage units summarized by $E(t)$ are given.
We refer to this problem as \emph{multi-period economic dispatch problem with storage} (MPED-S). The optimal value of this problem characterizes the total generation cost of meeting the loads when generators and storage charging/discharging are optimized. Denote the optimal value as a function of the storage locations and storage capacities by 
$J^{\mathrm{ED}}(\mathfrak{E}, \overline{\mathbf{s}})$. 

\begin{remark}[Stationary storage]\label{re:ss}
	It is easy to incorporate stationary storage into our model by disallowing relocation of a subset of mobile storage units. In other words, if storage unit $\widetilde{k}$ is a stationary storage, we require $i_{\widetilde{k}}(t)$ or the $\widetilde{k}$-th row of matrix $E(t)$ to stay constant over time. Alternatively, stationary storage can be modeled using a separate set of variables and constraints in the MPED-S problem (cf. \cite{qin2018submodularity}).
\end{remark}

The problem of optimizing the movements of the mobile storage fleet is then 
	\begin{align}\label{eq:om}
		\min_{\mathfrak{I}} \quad J^{\mathrm{ED}}(\mathfrak{E}(\mathfrak{I}), \overline{\mathbf{s}}) + J^{\mathrm{R}}(\mathfrak{I}),
	\end{align}
where the initial locations of the mobile storage units $\{i_k(1)\}_{k\in [K]}$ are given, and the decision variable $\mathfrak{I} := \{i_k(t)\}_{k\in [K], t\in [T]}$ is optimized such that each $i_k(t)\in \mathcal{I}_k^{\mathrm{S}}\subseteq [n]$ and $\mathcal{I}^{\mathrm{S}}_k$ is the set of buses with which mobile storage unit $k$ can be connected. 

Note that unlike the multi-period economic dispatch problem~\eqref{eq:ED} which is a convex program, the optimal storage relocation problem~\eqref{eq:om} is a combinatorial optimization problem. 



\section{Rapid Mobile Storage}\label{sec:fast}

We first consider storage that can both charge and move rapidly: it does not require any time to move between buses, and the power limit for charging and discharging is large enough that the power constraints \eqref{eq:ED-power} are never binding. This analysis serves two purposes: (a) it provides a theoretical account of the problem focusing on the \emph{relocation} aspect while omitting the complexity associated with travel times (which is addressed in Section \ref{sec:general}), and (b) it offers a good model for \emph{virtual mobile storage}, including storage capabilities derived from flexible loads that are geographically shiftable such as a collection of Internet data centers.  Given the locations of the mobile storage units $\mathfrak{E}$, the MPED-S problem is given by (\ref{eq:ED})  without the power constraint  \eqref{eq:ED-power}:
\begin{subequations} \label{eq:fastED} \noeqref{eq:fastED-obj}
	\begin{align} 
		\min_{\mathbf{g}, \mathbf{u}} \quad & \sum_{t=1}^T C_t (\mathbf{p}(t)) \label{eq:fastED-obj}\\
		\mbox{s.t.} \quad & \mathbf{1}^\top(\mathbf{g}(t) - \mathbf{d}(t) - E(t)\mathbf{u}(t)) = 0,& \!\!\!\!t\in [T], \label{eq:fastED:c1}\\ 
		&   H (\mathbf{g}(t) - \mathbf{d}(t) - E(t)\mathbf{u}(t)) \le \overline{\mathbf{f}}, & \!\!\!\!t\in [T],\label{eq:fastED:c2}\\
		&   \mathbf{0} \le L \mathbf{u}_k \le \overline{s}_k \mathbf{1}, &\!\!\!\! k \in [K]. \label{eq:fastED:c3}
	\end{align}
\end{subequations}
While we do not model the travel times in this case, we can still capture the cost of moving storage from one bus to another. In particular, we use $\kappa D_{ij}$ to model the cost of relocating from bus $i$ to bus $j$. For the example of Internet data centers, this cost may model the loss of quality of service from moving energy consuming data-processing loads from one data center to another. 
With relocation cost defined as \eqref{eq:fcost}, the optimal relocation problem still takes the form of \eqref{eq:om}. 

\subsection{Marginal value of rapid mobile storage}\label{sec:mv:fast}
Tackling the optimal relocation problem requires a good understanding of the optimal value of MPED-S problem~\eqref{eq:fastED}, which in turn depends on the spatial and temporal distribution of storage capacities. 
One way to characterize such dependence is through analyzing the \emph{marginal value of mobile storage} with fixed relocation pattern $\mathfrak{E}$.
\begin{definition}[Marginal value of mobile storage] 
	Given storage relocation pattern $\mathfrak{E}$, the marginal value of mobile storage $k$ is defined as
	\begin{equation}\label{eq:mv}
		\textsf{MV}^\mathrm{ms}_k(\mathfrak{E}, \overline{\mathbf{s}}) = - \frac{\partial J^{\mathrm{ED}}(\mathfrak{E}, \overline{\mathbf{s}})}{\partial \overline{s}_k}, \quad k \in [K].  
	\end{equation}
\end{definition}

$\textsf{MV}^\mathrm{ms}_k(\mathfrak{E}, \overline{\mathbf{s}})$ characterizes the reduction in the operation cost of the grid when we marginally increase the storage capacity of mobile storage $k$, given the relocation pattern $\mathfrak{E}$. As the objective function of \eqref{eq:fastED} is strictly convex quadratic, we can check that the partial derivatives in \eqref{eq:mv} indeed exist.	

It turns out that we can obtain an explicit and intuitive characterization of $\textsf{MV}^\mathrm{ms}_k(\mathfrak{E}, \overline{\mathbf{s}})$ via the dual variables of \eqref{eq:fastED}. Denote the (optimal) dual variables associated with constraints \eqref{eq:fastED:c1}, \eqref{eq:fastED:c2}, and \eqref{eq:fastED:c3} by $\gamma(t)\in \reals$, $\bm{\beta}(t)\in \reals^{2m}$, and $(\bm{\nu}_k, \bm{\mu}_k)\in \reals^{T}\times \reals^T$, respectively, where $ \bm{\nu}_k$ is associated with the lower bound in \eqref{eq:fastED:c3} and $\bm{\mu}_k$ is associated with the upper bound. We can also calculate the \emph{locational marginal prices} (LMPs), denoted by $\lambda_i(t)$,  for each bus $i$ and time period $t$ using these dual variables (cf. \cite{qin2018submodularity}):
\begin{equation}\label{eq:lmp}
	\bm{\lambda}(t) = \gamma(t) \ones - H^\top \bm{\beta}(t), \quad t \in [T]. 
\end{equation}
Notice that the dual variables and LMPs depend implicitly on the relocation pattern $\mathfrak{E}$ and storage capacities $\overline{\mathbf{s}}$. 


\begin{theorem}[Marginal value of  mobile storage]\label{th:mv-fast}
	The marginal value of mobile storage $k$ with  relocation $\mathfrak{E}$ is 
\begin{equation} \label{eq:mv-fast}
\textsf{MV}^\mathrm{ms}_k(\mathfrak{E}, \overline{\mathbf{s}})  = \ones^\top \bm{\mu}_k = \small\sum_{t=1}^T \Big( \lambda_{i_k(t+1)}(t+1) - \lambda_{i_k(t)}(t) \Big)_+,
\end{equation}
where $\bm{\lambda}(T+1):=\mathbf{0}$. 
\end{theorem}


The following observations are immediate. First, the marginal value of mobile storage is non-negative. In other words, increasing the capacities of mobile storage will weakly decrease the optimal cost for the dispatch problem. This is a consequence of the fact that $J^{\mathrm{ED}}(\mathfrak{E}, \overline{\mathbf{s}})$ is a convex and non-increasing function of $\overline{\mathbf{s}}$. Second, the marginal value of mobile storage can be directly obtained by summing up the dual variables associated with the upper bound in storage capacity constraints \eqref{eq:fastED:c3}. Indeed, the dual variable $\bm{\mu}_k$ characterizes the improvement in the objective function (generation costs) per unit relaxation of the constraint, i.e., per unit increase in the storage capacity. Finally, the marginal value of mobile storage $k$ can be calculated from the sum of (non-negative)  increases in LMPs along the relocation path of the mobile storage unit.


\subsection{Comparison to stationary storage and wires}\label{sec:econ}
Theorem~\ref{th:mv-fast} not only provides a way to relate the marginal value of a mobile storage unit to its relocation path, but also offers insights for understanding the value of mobile storage through well-understood quantities in electricity markets. In particular, we can compare the marginal value of a mobile storage to the marginal value of  wires and stationary energy storage. To this end, we denote the marginal value of the capacity associated with  $e$-th line capacity constraint by $\textsf{MV}^\mathrm{w}_e$, and denote the marginal value of  \emph{stationary energy storage} located at bus $i$ by $\textsf{MV}^\mathrm{ss}_i$. These quantities are related to the dual variables as follows \cite{qin2018submodularity, Taylor2015}:
\begin{equation}
	\textsf{MV}^\mathrm{w}_e = \sum_{t=1}^T  \beta_e(t), \quad \textsf{MV}^\mathrm{ss}_i = \sum_{t=1}^T\big(\lambda_i(t+1) - \lambda_i(t)\big)_+.
\end{equation}
For convenience, we also define the marginal values for each time period $t\in [T]$ as
\begin{equation}
	\textsf{MV}^\mathrm{w}_e(t) =  \beta_e(t), \quad \textsf{MV}^\mathrm{ss}_i(t) = \big(\lambda_i(t+1) - \lambda_i(t)\big)_+,
\end{equation}
\begin{equation}
	\textsf{MV}^\mathrm{ms}_k(t) = \mu_k(t) = \Big( \lambda_{i_k(t+1)}(t+1) - \lambda_{i_k(t)}(t) \Big)_+,
\end{equation}
for each $e\in [2m]$, $i\in [n]$ and $k\in [K]$, where we have omitted the dependence on $\mathfrak{E}$ and $\overline{\mathbf{s}}$ as they are fixed in this subsection.  

We provide three illustrative examples highlighting the role of mobile storage with different power network topologies and congestion patterns.

\subsubsection{Example 1: ``mobile storage = storage + wire''}
\begin{figure}[h]
\centering
\begin{subfigure}[b]{0.15\textwidth}
         \centering
         \includegraphics[width=\textwidth]{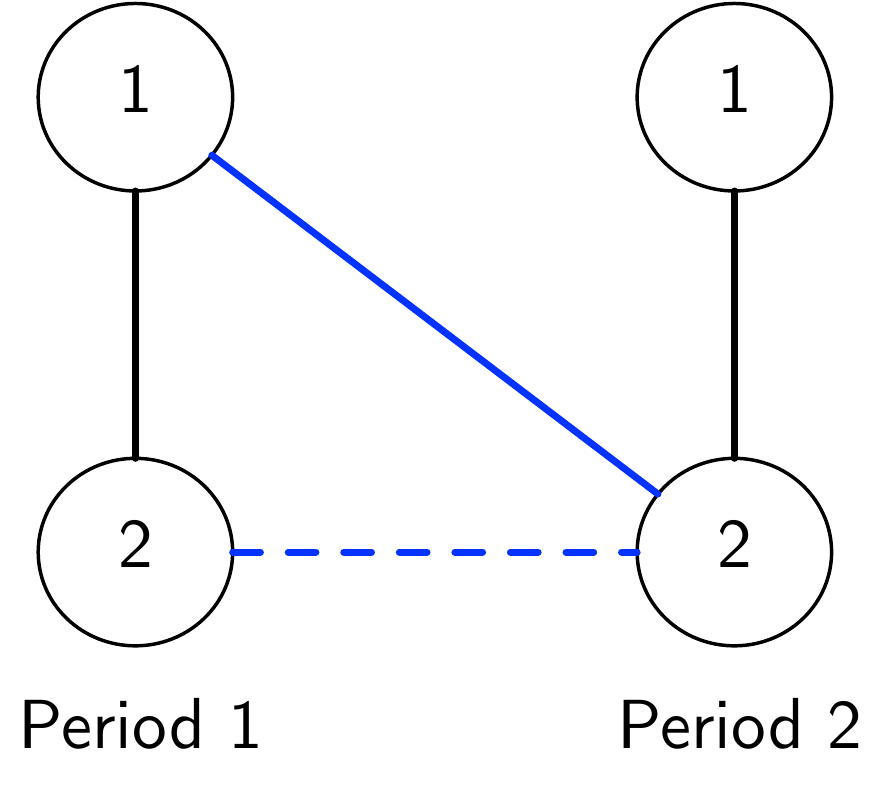}
         \caption{Example 1}
         \label{f1}
     \end{subfigure}
     \hfill
     \begin{subfigure}[b]{0.3\textwidth}
         \centering
         \includegraphics[width=\textwidth]{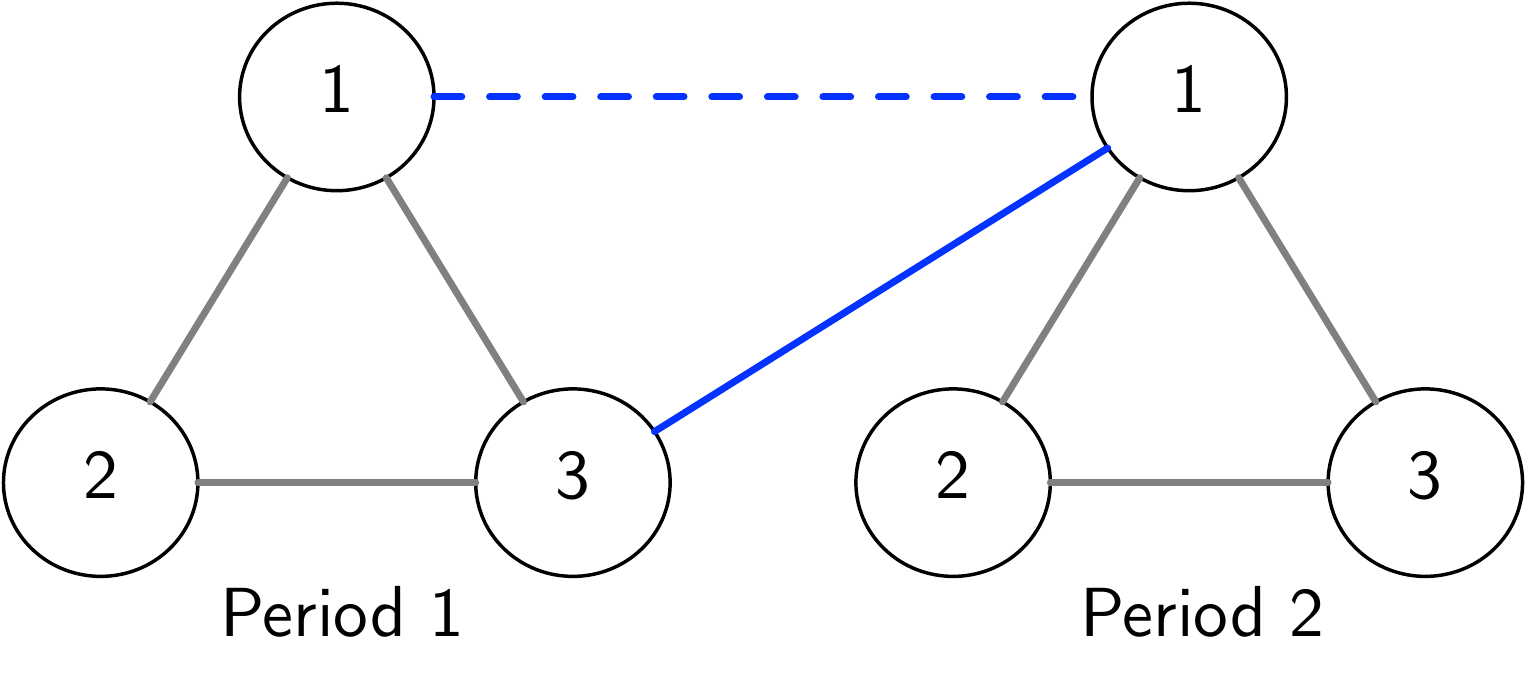}
         \caption{Example 2 and 3}
         \label{f2}
     \end{subfigure}
     \vspace{1ex}
     \caption{Time-extended graphs for Examples 1, 2 and 3.}
\end{figure}

Consider a two-bus network operated across two time periods. This is illustrated in Fig.~\ref{f1} using a time-extended graph of the network. Here each node in time period $t$ represents a bus in the power network, black solid lines represent electric wires, the blue solid line and the blue dashed line represent the ``power flow links'' created by a mobile storage unit moving from bus $1$ to bus $2$ and a stationary storage unit located at bus $2$, respectively. 

Suppose for this network we have $\lambda_2(2) > \lambda_2(1)> \lambda_1(1)$. Then we can obtain 
(see e.g., \cite{Taylor2015} for the relation between LMPs and dual variables associated with transmission constraints for a radial network) 
that 
	$\textsf{MV}^\mathrm{w}(1) = \lambda_2(1) - \lambda_1(1)$,  
  $\textsf{MV}^\mathrm{ss}(1) = \lambda_2(2) - \lambda_2(1)$, and $\textsf{MV}^\mathrm{ms}(1) = \lambda_2(2) - \lambda_1(1)$, where we have omitted the subscripts for the marginal values as they are clear from the context. As a result, 
   \begin{equation}
   	\textsf{MV}^\mathrm{ms}(1) = \textsf{MV}^\mathrm{w}(1) + \textsf{MV}^\mathrm{ss}(1),
   \end{equation}
   in this case. This is intuitive as both flow paths $(i=1, t=1) \rightarrow (i=2, t=2)$ and $(i=1,t=1) \rightarrow (i=2,t=1) \rightarrow (i=2,t=2)$ enable sending energy from $(i=1, t=1)$ to $(i=2, t=2)$. We can generalize this observation to other radial power networks, and make a statement about the value of mobile storage in a radial network where the power flow is along a directed flow path. 
\begin{lemma}[Radial network]\label{lem:radial}
	Consider a radial network where the power flow between node $i$ at time $t$ and node $j$ at time $t+1$ is along the directed flow path $(i,t) \rightarrow (k_1,t) \rightarrow ... \rightarrow (k_p,t) \rightarrow (j, t) \rightarrow (j, t+1)$ connecting them in the time-extended graph, where $k_1, ... k_p$ are nodes on the power network, and each edge represents a wire or a stationary storage unit. The marginal value of a mobile storage that moves from node $i$ at time $t$ to node $j$ at time $t+1$ is equal to the sum of marginal values of wires on the path from bus $i$ to bus $j$ on the power network for period $t$ and the marginal value of stationary storage located at bus $j$ for period $t$.
\end{lemma}
\subsubsection{Example 2: ``mobile storage $>$ storage + wire''}
Now we consider the operation of a three bus network across 2 periods as depicted in Fig.~\ref{f2}. The mobile storage unit moves from bus $3$ to bus $1$. Suppose that $\lambda_1(2)>\lambda_1(1) > \lambda_3(1)$, and  the line $3\rightarrow 1$ and another line in the loop (i.e., $3\rightarrow 2$ or $2\rightarrow 1$ in power network) are congested in period $1$. Here we have $\beta_{3\rightarrow 1}(1) < \lambda_1(1) - \lambda_1(3)$, where $\beta_{3\rightarrow 1}(1)$ is the dual variable associated with the line capacity constraint for the flow from bus $3$ to bus $1$ in time period $1$.   Appendix~\ref{app:ex2data} provides the data for a problem instance where this holds.
In this case, we have $\textsf{MV}^\mathrm{ms}(1) = \lambda_1(2) - \lambda_3(1)$, $\textsf{MV}^\mathrm{ss}(1) = \lambda_1(2) - \lambda_1(1)$, and $\textsf{MV}^\mathrm{w}(1) = \beta_{3\rightarrow 1}(1) < \lambda_1(1) - \lambda_1(3)$, and
\begin{equation}
	  	\textsf{MV}^\mathrm{ms}(1) > \textsf{MV}^\mathrm{w}(1) + \textsf{MV}^\mathrm{ss}(1),
\end{equation}
where $ \textsf{MV}^\mathrm{w}(1)$ denotes the marginal value of the line connecting bus $3$ and bus $1$ in time period $1$. The gap stems from the fact that the flow paths $(i=3,t=1) \rightarrow (i=1, t=2)$ and $(i=3,t=1) \rightarrow (i=1,t=1) \rightarrow (i=1, t=2)$ are not equivalent. This is because the flow on a line in a loop cannot be freely determined for power networks with loops due to Kirchhoff's voltage law. When another line in the loop is also congested, 
 increasing the capacity of link $3\rightarrow 1$ in the power network by one unit does not let us increase flow on the link due to the binding capacity constraint of the other congested line. This effect is referred to as \emph{loop externality} \cite{Wu1996}.

\subsubsection{Example 3: ``mobile storage $<$ storage + wire''}
 Consider the same network and storage configuration in Example 2 (Fig.~\ref{f2}). Now we require that $\lambda_1(2)>\lambda_1(1) >\lambda_2(1)> \lambda_3(1)$, and  the line $3\rightarrow 1$ is the only line congested in period $1$. In this case, increasing the capacity of line $3\rightarrow 1$ can increase not only the flow on the link $3\rightarrow 1$, but also allow the flow along the links $3\rightarrow 2\rightarrow 1$ to be increased. As a result, we can have $\beta_{3\rightarrow 1} > \lambda_1(1) - \lambda_3(1)$. It follows that in this case, we have 
\begin{equation}
	  	\textsf{MV}^\mathrm{ms}(1) < \textsf{MV}^\mathrm{w}(1) + \textsf{MV}^\mathrm{ss}(1),
\end{equation}
where $ \textsf{MV}^\mathrm{w}(1)$ denotes the marginal value of the line connecting bus $3$ and bus $1$ in time period $1$. Appendix~\ref{app:ex2data} provides a numerical example where this holds.


\subsection{Optimal relocation of small mobile storage}

Solving the optimal relocation problem for rapid mobile storage using brute-force methods is computationally intractable because the computation time scales with the problem size as $O(n^{KT})$. However, in practice, we can simplify the optimal relocation problem for a mobile storage fleet because we expect that the storage capacities of such fleets will be relatively 
 \emph{small} in the near future. This is motivated by the fact that 
 the total storage capacity is still a tiny fraction of the total power load. 
 To get a sense, the estimated total capacity of storage deployed in 2020  is about $0.296\%$ of the estimated load in the same year \cite{GTMstorage20}. We formalize this concept in the following assumption:
 \begin{assumption}[Small storage]\label{ass:small}
 	We assume that the storage capacities $\overline{\mathbf{s}}$ are small in either of the following two senses:
\begin{enumerate}[{\bf 
A1}.a]
      \item Then $\arg\min_{\mathfrak{I}} \{J^{\mathrm{ED}}(\mathfrak{E}(\mathfrak{I}), \overline{\mathbf{s}}) + J^{\mathrm{R}}(\mathfrak{I})\} = \arg\min_{\mathfrak{I}} \{\widehat{J}^{\mathrm{ED}}(\mathfrak{E}(\mathfrak{I}), \overline{\mathbf{s}}) + J^{\mathrm{R}}(\mathfrak{I})\}$, where $\widehat{J}^{\mathrm{ED}}(\mathfrak{E}, \overline{\mathbf{s}})$ is the first order Taylor approximation of $J^{\mathrm{ED}}(\mathfrak{E}, \overline{\mathbf{s}})$:
 	$
 		\widehat{J}^{\mathrm{ED}}(\mathfrak{E}, \overline{\mathbf{s}}) = J^{\mathrm{ED}}(\mathfrak{E}, \mathbf{0}) -
 		\sum_{k=1}^K \textsf{MV}^\mathrm{ms}_k(\mathfrak{E}, \mathbf{0}) \overline{s}_k. 
 	$

  \item Suppose the load is such that the LMPs for each node and time with $\overline{\mathbf{s}} = 0$ are distinct. Then a small capacity $\overline{\mathbf{s}}$ does not change the relative ordering of LMPs. \footnote{This holds for almost every load $\mathbf{d}(t) \in \mathbb{R}^n$ for each $t$. In other words, the set of loads for which this does not hold has Lebesgue measure 0.}
  \end{enumerate}
  
 \end{assumption}
 Note that Assumption 1 does not rule out the possibility that there is a significant amount of \emph{stationary storage} connected to the system. Stationary storage can be modeled using a separate set of variables and constraints in the MPED-S problem. 

Under the small storage assumption, the optimal relocation problem for the fleet decouples. In other words, solving the optimal relocation problem for a fleet is equivalent to solving $K$ optimal relocation problems for individual units: 
\begin{equation}\label{eq:omk}
	\max_{\mathfrak{I}_k} \quad \textsf{MV}^\mathrm{ms}_k(\mathfrak{E}, \mathbf{0}) \overline{s}_k - J_k^{\mathrm{R}}(\mathfrak{I}_k),\quad k \in [K],
\end{equation}
where the initial locations of the mobile storage units $\{i_k(1)\}_{k\in [K]}$ are given, and the decision variable $\mathfrak{I}_k= \{i_k(t)\}_{ t\in [T]}$ is optimized such that each $i_k(t)\in \mathcal{I}_k^{\mathrm{S}}\subseteq [n]$ and $\mathcal{I}^{\mathrm{S}}_k$ is the set of buses with which mobile storage unit $k$ can be connected.  The optimal trajectory for different storage units may be different because they have different storage capacities, initial locations and sets of admissible buses $\mathcal{I}^{\mathrm{S}}_k$.

\begin{figure}[h]
\centering
\includegraphics[width=.4\textwidth]{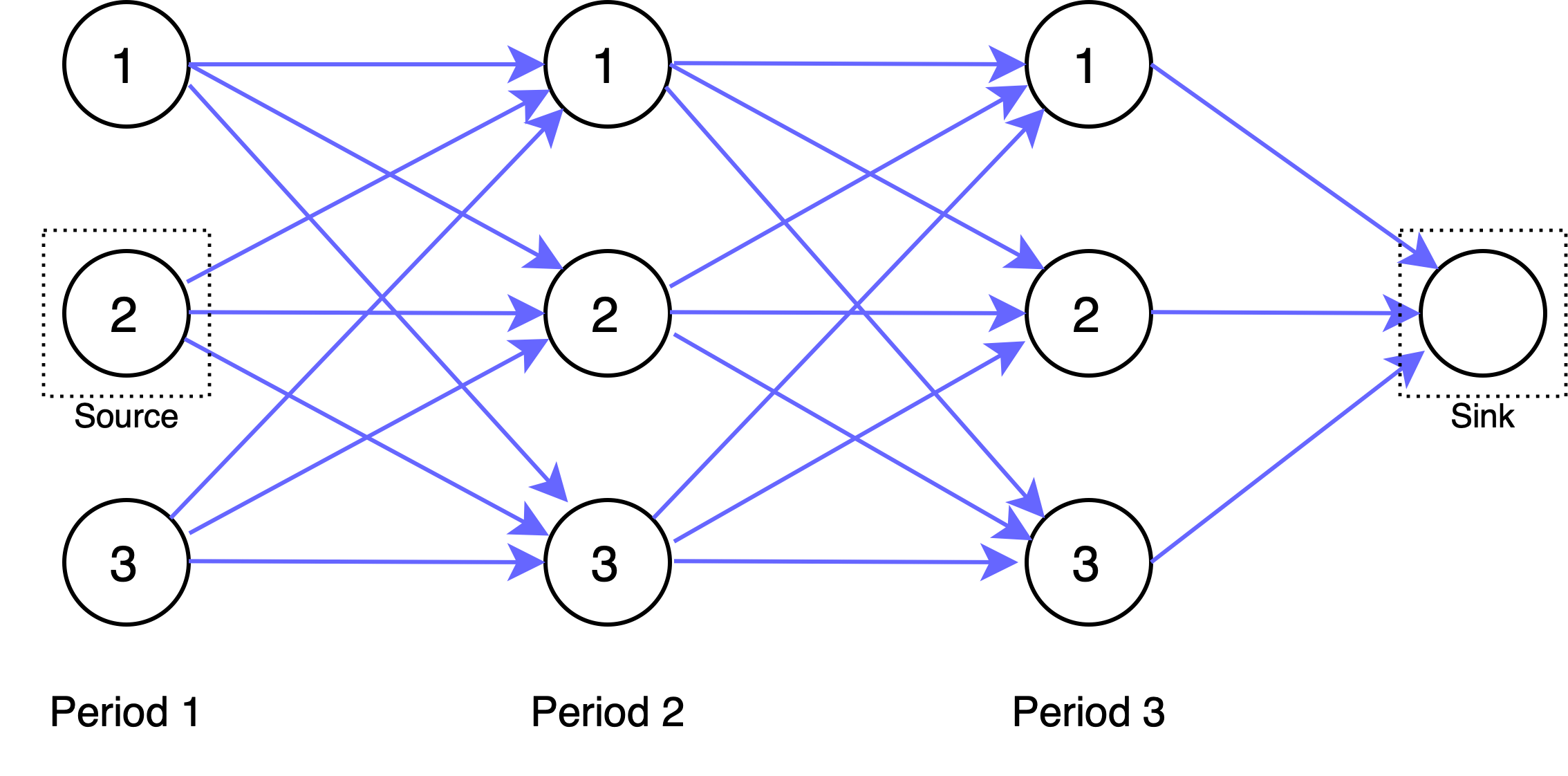}
\caption{Example of the time extended graph on which the shortest path problem is defined. In this example, $n=3$, $T=3$, and the initial location for storage $k$ is $i_k(1)=2$. }\label{f3}	
\end{figure}

The optimal relocation problem~\eqref{eq:omk} can be converted into a shortest path problem and thus solved efficiently in polynomial time ($O(n^2 T)$) with a range of algorithms including the Bellman-Ford algorithm and linear programming. Indeed, we can construct a time-extended graph $G(\mathcal{V}, \mathcal{E})$, where the set of nodes $\mathcal{V}$ includes $T$ copies of all the nodes in the power network $[n]$, and a dummy sink node. The set of edges $\mathcal{E}$ includes \emph{directed} edges from every node $i$ in the $t$-th copy to every node $j$ in the $(t+1)$-th copy, with edge weight
\begin{equation}
	w_{ij}(t) = \kappa D_{ij} - \overline{s}_k (\lambda_j(t+1) - \lambda_i(t))_+, \quad t\in [T-1],
\end{equation}
 and \emph{directed} edges from every node in the $T$-th copy to the dummy sink node, with edge weight $0$. The LMPs are calculated  using dual variables for \eqref{eq:fastED} with $\overline{\mathbf{s}}=\mathbf{0}$. By solving the shortest path problem from source node $i_k(1)$ to the dummy sink node in the time extended graph, we can identify the optimal trajectory $\{i_k(t)\}_{t\in [T]}$ for storage $k$.  
Fig.~\ref{f3} provides an example of the graph $G(\mathcal{V}, \mathcal{E})$.

%

\section{General Mobile Storage}\label{sec:general}
In general, the time needed for relocating mobile storage units is not negligible, and both energy limits and power limits should be considered for mobile storage operation. In this case, we need to solve the general version of the MPED-S problem	 \eqref{eq:ED} and the optimal relocation problem \eqref{eq:om}. 



\subsection{Marginal value of mobile storage}\label{sec:gen:mvms}

Similar to Section~\ref{sec:mv:fast}, we quantify the marginal value of mobile storage for a fixed relocation pattern $\mathfrak{E}$. 
 Denote the (optimal) dual variables associated with constraints \eqref{eq:ED-demand} and \eqref{eq:ED-line} by $\gamma(t)\in \reals$ and $\bm{\beta}(t)\in \reals^{2m}$. $ \bm{\nu}_k$ and $\bm{\mu}_k$ are associated with the lower bound and upper bound in \eqref{eq:ED-energy}, and $ \bm{\omega}_k$ and $\bm{\phi}_k$ are associated with the lower bound and upper bound in \eqref{eq:ED-power}, respectively. We define the LMPs $\bm{\lambda}(t)$ as in \eqref{eq:lmp}.
 
 When the power limits depend on the energy limits (i.e., $\bar u_k$ is a function of $\bar s_k$), it is necessary for us to keep track of whether energy or power constraints are binding at each time. For simplicity, we introduce the 
 the following standard assumption for the MPED-S problem under consideration.
\begin{assumption}[LICQ]\label{ass:licq}
Given any relocation pattern $\mathfrak{E}$, the constraints binding at the solution of the MPED-S problem \eqref{eq:ED} are linearly independent.
\end{assumption}

When optimizing the movement of mobile storage in a power network, we may be able to forecast the LMPs ahead of time. These prices will not be affected by the movement of mobile storage under the small storage assumption (Assumption \ref{ass:small}). Meanwhile, we can identify the set of binding storage constraints for each storage $k$ using a simpler optimization involving LMPs and the parameters for the storage unit.

\begin{lemma}\label{lem:arbitrage}
Suppose that for each storage $k$, the price arbitrage problem given by 
\begin{equation}\label{eq:arbitrage}
	\begin{aligned}
	\max_{\mathbf{u}_k} \;  &-\sum_{t=1}^T \lambda_{i_k(t)}(t) u_k(t)\\
		\mbox{s.t.} \; &   \mathbf{0} \le L \mathbf{u}_k \le \overline{s}_k \mathbf{1};  -  \overline{u}_k(\overline{s}_k)\bm{\Delta}^{\mathrm{S}}_k  \le \ub_k \le \overline{u}_k(\overline{s}_k)\bm{\Delta}^{\mathrm{S}}_k.
	\end{aligned}
\end{equation}
has a unique solution. Then the optimal operation $\mathbf{u}_k$ for storage $k$ in a solution of the MPED-S problem \eqref{eq:ED} is the same as the solution of the price arbitrage problem \eqref{eq:arbitrage}.
\end{lemma}


\begin{corollary}\label{cor:order-of-constraints}
Since the optimal operation of storage $k$ in the MPED-S problem \eqref{eq:ED} coincides with the solution of the price arbitrage problem \eqref{eq:arbitrage}, they will lead to the same collection of binding constraints for storage $k$. 
\end{corollary}

\begin{figure}[h] 
\centering
\begin{tikzpicture}	
\begin{axis}
       [
         axis x line*  = bottom,
         axis y line  = left,
         hide y axis,
         width        = .4\textwidth,
         height       = .12\textwidth,          
         ymax         = 1,
         ymin         = 0,
         ytick        = {0,1},
         xmax         = 7,
         xmin         = 0,
         xtick        = {1,2,3,4,5,6},
         xticklabel shift={20pt},
        xticklabels={$t^\mathrm{p}_1$, $t^\mathrm{p}_2$, $t^\mathrm{e}_1$, $t^\mathrm{p}_3$,  $t^\mathrm{e}_2$,\qquad\qquad Time },
        xlabel style={yshift=-0.2cm},
        x tick label style={
            rotate=0,
            anchor=south,
        },
       ]
       \node[anchor=north] (s) at (axis cs:1,0){};
       \node[anchor=north] (ss) at (axis cs:2,0){};
       \node (d) at (axis cs:3,0){};
       \draw[->,>=stealth](s) to [out=90,in=90] (d);
       \draw[->,>=stealth](ss) to [out=90,in=90] (d);

       \node[anchor=north] (sss) at (axis cs:4,0){};
       \node (dd) at (axis cs:5,0){};
       \draw[->,>=stealth](sss) to [out=90,in=90] (dd);

      \end{axis}

\end{tikzpicture}
\caption{$\sigma$ operator} \label{fig:sigma}
\end{figure}
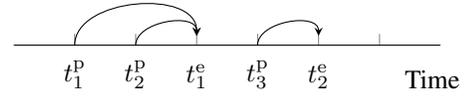

Under Assumption \ref{ass:licq}, the binding constraints in \eqref{eq:arbitrage} can then be used to partition $[T]$ into two disjoint sets: 
\[
\mathcal{T}_k^\mathrm{e} (\mathfrak{E}) = \{t_r^\mathrm{e}\}_{r\in [T^\mathrm{e}_k]}, \quad 
\mathcal{T}_k^\mathrm{p}(\mathfrak{E}) = \{t_s^\mathrm{p}\}_{s\in [T^\mathrm{p}_k]}
\]
where $\mathcal{T}_k^\mathrm{e}(\mathfrak{E})$ and $\mathcal{T}_k^\mathrm{p}(\mathfrak{E})$ represent the times when energy capacity and power capacity constraints are binding, and $T_k^\mathrm{e}$ and $T_k^\mathrm{p}$ are the number of time periods within $\mathcal{T}_k^\mathrm{e}(\mathfrak{E}), \mathcal{T}_k^\mathrm{p}(\mathfrak{E})$, respectively. Henceforth we omit the dependence on $\mathfrak{E}$ and $k$ to simplify the notation. Let $\sigma: \mathcal{T}^\mathrm{p}\mapsto \mathcal{T}^\mathrm{e}$ denote the mapping from
 each power capacity constrained time to the next energy capacity constrained time, i.e., 
$
\sigma(t^\mathrm{p}) = \inf_{\tau} \{\tau \in \mathcal{T}^\mathrm{e}: \tau > t^\mathrm{p}\}.
$
Fig. \ref{fig:sigma} provides an illustration of $\sigma$. 
The $\sigma$ mapping is defined for all $t^\mathrm{p}\in \mathcal{T}^\mathrm{p}$ when $T \in \mathcal{T}^\mathrm{e}$. This is  the case when we have nonnegative LMPs since the optimal state of charge (SoC) in the terminal time period $T$ will be empty. 

\begin{theorem}[Marginal value of mobile storage]\label{th:mv-general}
	The marginal value of mobile storage $k$  with relocation $\mathfrak{E}$ is
\begin{equation}\label{eq:mv-general}
\begin{split}
 &\textsf{MV}^\mathrm{ms}_k(\mathfrak{E}, \mathbf{\bar{s}}) =  \mathbf{1}^\top \bm{\mu}_k + \bar{u}_k'(\bar{s}_k) (\bm{\Delta}^{\mathrm{S} }_{k})^\top (\bm{\omega}_k + \bm{\phi}_k) ,
 \end{split}
\end{equation}
where for each $t^\mathrm{e}_{r} \in \mathcal{T}^\mathrm{e}$,
\begin{align}
&\mu_k(t^\mathrm{e}_{r}) = \left( \lambda_{i_k (t^\mathrm{e}_{r+1})}(t^\mathrm{e}_{r+1}) - \lambda_{ i_k(t^\mathrm{e}_{r})} (t^\mathrm{e}_{r}) \right)_+, \quad\quad\quad\,\,\,\\
&\omega_k(t^\mathrm{e}_r) + \phi_k(t^\mathrm{e}_r) = 0,
\end{align}
 for each $t^\mathrm{p}_{s} \in \mathcal{T}^\mathrm{p}$,
\begin{align}
&\mu_k(t^\mathrm{p}_s)= 0,\\
&\omega_k(t^\mathrm{p}_s) + \phi_k(t^\mathrm{p}_s) = \left| \lambda_{i_k(\sigma(t^\mathrm{p}_{s}))} (\sigma (t^\mathrm{p}_{s})) - \lambda_{i_k(t^\mathrm{p}_{s})}( t^\mathrm{p}_{s}) \right|,
\end{align}
and we define $t^\mathrm{e}_{T^\mathrm{e}+1}:=T+1$ and $\bm{\lambda}(T+1) := 0$.
\end{theorem}



Similar to the rapid storage case, the marginal value of mobile storage is non-negative, i.e., increasing the storage capacity $\bar{s}_k$ will weakly decrease the optimal cost for the dispatch problem. The marginal value is obtained by summing the dual variable associated with the upper bound in the storage capacity constraint \eqref{eq:ED-energy}, and the dual variables associated with the power constraints \eqref{eq:ED-power} weighted by the time available to charge $\bm{\Delta}^{\mathrm{S} }_{k}$ and the dependence of power capacity on storage capacity $\bar{u}_k'(\bar{s}_k)$. 
The dual variable $\bm{\omega}_k$ characterizes the value of increasing the discharging power capacity, and the dual variable $\bm{\phi}_k$ characterizes the value of increasing the charging power capacity. 
Comparing the marginal value expression in \eqref{eq:mv-general} to that of rapid mobile storage \eqref{eq:mv-fast}, we observe that the new expression uses LMP increases across consecutive energy capacity-constrained time periods as well as LMP differences across power constrained and energy constrained time periods. The price arbitrage across $t^\mathrm{e}_{r} \in \mathcal{T}^\mathrm{e}$ is not a complete measure of the marginal value, as charge/discharge also occurs at intermediate time steps ($t^\mathrm{p}_{r} \in \mathcal{T}^\mathrm{p}$).

\begin{corollary}
In the case that power constraints have no dependence on $\overline{s}_k$, i.e. $\overline{u}_k'(\overline{s}_k) = 0$, the marginal value is 
\begin{equation}
\begin{split}
 \textsf{MV}^\mathrm{ms}_k(\mathfrak{E}, \mathbf{\bar{s}}) =   \mathbf{1}^\top \bm{\mu}_k
 \end{split}
\end{equation}
with $\mu_k(t)$ defined as in Theorem \ref{th:mv-general}.
\end{corollary}

This is similar to the case in Theorem \ref{th:mv-fast}, as any increase in storage capacity will not alleviate power constraints. However, the expression for $\bm{\mu}_k$ here is different from that in Theorem \ref{th:mv-fast}  due to the storage power constraints.
Stationary storage can also be viewed as a special case of mobile storage where the battery is located at a single bus at all times. 
\begin{corollary}[Marginal value of stationary storage]
The marginal value of stationary storage $k$ located at bus $i$, i.e., $i_k(t) = i$ for all $t$, is
\begin{equation}\label{eq:mv-ss}
\begin{split}
 &\textsf{MV}^\mathrm{ss}_k(\mathbf{\bar{s}}) =  \mathbf{1}^\top \bm{\mu}_k + \bar{u}_k'(\bar{s}_k) \Delta \mathbf{1}^\top (\bm{\omega}_k + \bm{\phi}_k) 
 \end{split}
\end{equation}
where for each $t^\mathrm{e}_{r} \in \mathcal{T}^\mathrm{e}$,
\begin{align}
&\mu_k(t^\mathrm{e}_{r}) = \left( \lambda_{i}(t^\mathrm{e}_{r+1}) - \lambda_{ i} (t^\mathrm{e}_{r}) \right)_+, \,\,\omega_k(t^\mathrm{e}_r) + \phi_k(t^\mathrm{e}_r) = 0,
\end{align}
 for each $t^\mathrm{p}_{s} \in \mathcal{T}^\mathrm{p}$,
\begin{align}
&\mu_k(t^\mathrm{p}_s)= 0,\,\, \omega_k(t^\mathrm{p}_s) + \phi_k(t^\mathrm{p}_s) = \left| \lambda_{i} (\sigma (t^\mathrm{p}_{s})) - \lambda_{i}( t^\mathrm{p}_{s}) \right|,
\end{align}
and  we define $t^\mathrm{e}_{T^\mathrm{e}+1}:=T+1$ and $\bm{\lambda}(T+1) := 0$.
\end{corollary}

This result generalizes existing results  on the locational marginal value of stationary storage \cite{bose2016marginal, qin2018submodularity} by incorporating storage power constraints.

\subsection{Comparison to stationary storage and wires}\label{sec:econ-general}
Since the rapid mobile storage model discussed in Section~\ref{sec:fast} is a special case of the general mobile storage model, the examples discussed in Section~\ref{sec:econ} remain valid for the general mobile storage model. However, Lemma~\ref{lem:radial} does not generalize to the general mobile storage case as 
the storage power constraints and time taken to travel can impact the marginal value of mobile storage. Intuitively, with a nonzero travel time, the time available for mobile storage to charge/discharge is strictly less than that for stationary storage. As a result, in scenarios where the storage power constraints are binding and the reduced charging/discharging time matters, the marginal value of mobile storage can be strictly less than the sum of marginal values of the corresponding stationary storage unit and wires even if the network does not have loops.


\subsection{Optimal relocation of small mobile storage}\label{subsec:relocation-general}

Under Assumption~\ref{ass:small}, the optimal relocation problem for the mobile storage fleet decouples as in the case of rapid mobile storage, and we can use LMPs for $\overline{\mathbf{s}}=\mathbf{0}$ in the marginal value calculation. 
However, the marginal value of general mobile storage depends on the order of binding constraints.
When the constraint binding pattern $(\mathcal{T}^\mathrm{e}, \mathcal{T}^\mathrm{p})$ is given, we can identify the optimal relocation associated with $(\mathcal{T}^\mathrm{e}, \mathcal{T}^\mathrm{p})$ by solving a shortest path problem referred to as \textsf{SP-E}$(\mathcal{T}^\mathrm{e}, \mathcal{T}^\mathrm{p})$ over all energy constrained time periods.
 We construct a time-extended graph $G(\mathcal{V}, \mathcal{E})$ where  the set of nodes $\mathcal{V}$ includes $T^\mathrm{e}$ copies of all the nodes in the power network $[n]$, and a dummy sink node. The set of edges $\mathcal{E}$ includes \emph{directed} edges from every node $i$ in the  copy corresponding to $t^\mathrm{e}_r$ to every node $j$ in the copy corresponding to $t^\mathrm{e}_{r+1}$, with an edge weight
\begin{align}
&w_{ij}(t^\mathrm{e}_r)= \label{eq:ew}\\
&\begin{cases}
 - \overline{s}_k \left( \lambda_{j}(t^\mathrm{e}_{r+1}) - \lambda_{ i} (t^\mathrm{e}_{r}) \right)_++ \kappa D_{ij}, & \mbox{if }t^\mathrm{e}_{r+1} = t^\mathrm{e}_{r}+1,\\
- \overline{s}_k \left( \lambda_{j}(t^\mathrm{e}_{r+1}) - \lambda_{ i} (t^\mathrm{e}_{r}) \right)_+ + J^\mathsf{SP\mbox{-}P}_{ij}(t^\mathrm{e}_{r}),	& \mbox{otherwise}, \nonumber
\end{cases}
\end{align}
and \emph{directed} edges from every node in the $T^\mathrm{e}$-th copy to the dummy sink node, with edge weight $0$. If $t^\mathrm{e}_{r+1} \neq t^\mathrm{e}_{r}+1$, there are power constrained time periods between $t^\mathrm{e}_{r}$ and $t^\mathrm{e}_{r+1}$, and the edge weight will depend on the cost of traveling along an optimal path for all intermediate power constrained time periods $t^\mathrm{p}_s \in \{t^\mathrm{e}_r+1,\dots,  t^\mathrm{e}_{r+1}-1\}$ denoted by $J^\mathsf{SP\mbox{-}P}_{ij}(t^\mathrm{e}_{r})$.

\subsubsection{Calculating $J^\mathsf{SP\mbox{-}P}_{ij}(t^\mathrm{e}_{r})$}
The optimal path for the power constrained time periods between $t^\mathrm{e}_{r}$ and $t^\mathrm{e}_{r+1}$ can be found by 
 formulating a shortest path problem referred to as \textsf{SP-P}$_{ij}(t^\mathrm{e}_{r})$. For each $i,j\in [n]$,  $t^\mathrm{e}_{r}\in \Tcal^\mathrm{e}$, we construct a shortest path problem on a time-extended graph $\widetilde{G}(\mathcal{\widetilde V}, \mathcal{\widetilde E})$ where we omit the dependence on $i$, $j$ and $t^\mathrm{e}_{r}$ to simplify notation. The set of nodes $\mathcal{\widetilde V}$ include a source node (representing node $i$ in time $t_r^\mathrm{e}$), a sink node (representing node $j$ in time $t_{r+1}^\mathrm{e}$), and a copy of all the nodes in the power network $[n]$ for each $t^\mathrm{p}_s \in \{t^\mathrm{e}_r+1,\dots,  t^\mathrm{e}_{r+1}-1\}$. The set of edges $\mathcal{\widetilde E}$ includes \emph{(a)} a directed edge from the source node $i$ to every node $\widetilde{j}\in [n]$ in the first copy (which corresponds to $t^\mathrm{p}_s =t^\mathrm{e}_r+1$), with weight
 $
 	w_{i\,\widetilde{j}} = \kappa D_{i\,\widetilde{j}},
$
 \emph{(b)} if $|\{t^\mathrm{e}_r+1,\dots,  t^\mathrm{e}_{r+1}-1\}|>1$, a directed edge from every node $\widetilde i\in [n]$ in the copy corresponding to each $t^\mathrm{p}_s\in \{t^\mathrm{e}_r+1,\dots,  t^\mathrm{e}_{r+1}-2\}$ to every node $\widetilde j \in [n]$ in the copy corresponding to $t^\mathrm{p}_{s+1}$, with weight
\begin{equation}\label{eq:ewp}
w_{\,\widetilde{i}\,\, \widetilde{j}}(t^\mathrm{p}_s)  = \kappa D_{\,\widetilde{i}\,\, \widetilde{j}} - \overline{s}_k \overline{u}_k'(\overline{s}_k) \Delta^\mathrm{S}_{\,\widetilde{i}\,\, \widetilde{j}}  | \lambda_j(t^\mathrm{e}_{r+1}) - \lambda_{\,\widetilde{i}}(t^\mathrm{p}_{s}) |,
\end{equation}
and \emph{(c)} a directed edge from each node $\widetilde i \in [n]$ in the last copy (corresponding to $t^\mathrm{p}_s =t^\mathrm{e}_{r+1}-1$) to the sink node $j$, with  weight defined in \eqref{eq:ewp}. We overload the notation $\Delta^\mathrm{S}_{ij}$ to represent the time available for battery operation if it travels from node $i$ to $j$ in that time period. $J^\mathsf{SP\mbox{-}P}_{ij}(t^\mathrm{e}_{r})$ is the cost of traveling along the shortest path in this extended graph. 

\subsubsection{Solving \textsf{SP-E}$(\Tcal^\mathrm{e},\Tcal^\mathrm{p})$}
In order to solve \textsf{SP-E}$(\Tcal^\mathrm{e},\Tcal^\mathrm{p})$, we first solve $J^\mathsf{SP\mbox{-}P}_{ij}(t^\mathrm{e}_{r})$ for each $i,j\in [n]$,  $t^\mathrm{e}_{r}\in \Tcal^\mathrm{e}$ and determine the edge weights of the time-extended graph $G(\mathcal{V}, \mathcal{E})$. We then solve the shortest path problem \textsf{SP-E}$(\Tcal^\mathrm{e},\Tcal^\mathrm{p})$ using the edge weights determined by \eqref{eq:ew}.

\subsubsection{Identifying optimal relocation $\mathfrak{E}$}
First, from the solution of \textsf{SP-E}$(\Tcal^\mathrm{e},\Tcal^\mathrm{p})$, we obtain the sequence of nodes where mobile storage is located at each time in $\Tcal^\mathrm{e}$. We use this to obtain the nodes $(i,j)$ at which the mobile storage is located for each pair of non-consecutive energy constrained time periods ($t^\mathrm{e}_{r}, t^\mathrm{e}_{r+1}$ such that $t^\mathrm{e}_{r+1} \neq t^\mathrm{e}_{r}+1$), and the solution to \textsf{SP-P}$_{ij}(t^\mathrm{e}_r)$ provides the relocation pattern for the storage during the power constrained time periods between $t^\mathrm{e}_r$ and $t^\mathrm{e}_{r+1}$. Piecing together the locations of the storage in energy constrained times and power constrained times results in the optimal relocation pattern $\mathfrak{E}(\Tcal^\mathrm{e},\Tcal^\mathrm{p})$ given $(\Tcal^\mathrm{e},\Tcal^\mathrm{p})$.

In practice, we do not know the constraint binding patterns $(\mathcal{T}^\mathrm{e}, \mathcal{T}^\mathrm{p})$ without knowing the optimal relocation, as moving along different relocation paths will change the optimal storage operation. One approach is to enumerate all possible constraint binding patterns $(\mathcal{T}^\mathrm{e}, \mathcal{T}^\mathrm{p})$ and solve \textsf{SP-E}$(\Tcal^\mathrm{e},\Tcal^\mathrm{p})$ for each constraint binding pattern. This will result in an optimal relocation pattern $\mathfrak{E}(\Tcal^\mathrm{e},\Tcal^\mathrm{p})$, whose consistency with the assumed constraint binding pattern needs to be checked by computing the actual constraint binding pattern under this relocation pattern and comparing with the assumed $(\Tcal^\mathrm{e},\Tcal^\mathrm{p})$. If they are the same, this relocation pattern is an admissible solution. After going through all $O(2^T)$ possibilities, we can find the optimal relocation pattern by finding the admissible solution with the lowest cost.
Since for small storage either energy or power constraint is binding at every time step, this amounts to solving the shortest path problem described above $O(2^T)$ times. As our model focuses on daily operation with every time period long enough for mobile storage to relocate, $T$ is usually relatively small (e.g., $T=12$ for $2$ hour time periods). 
 Algorithm \ref{alg:relocation} summarizes the steps to generate an optimal relocation path for general mobile storage.
 
 \SetKwInOut{Output}{Output}
  \SetKwInOut{Input}{Input}
  
 \begin{algorithm}[h]
\SetAlgoLined

 \For{\emph{each possible} $(\mathcal{T}^\mathrm{e}, \mathcal{T}^\mathrm{p})$ }{
\For{$t^\mathrm{e}_r \in \mathcal{T}^\mathrm{e}$, $i,j \in [n]$} {
Compute $J^\mathsf{SP\mbox{-}P}_{ij}(t^\mathrm{e}_{r})$ by solving \textsf{SP-P}$_{ij}(t^\mathrm{e}_r)$;\\
Compute  weights in \textsf{SP-E}$(\mathcal{T}^\mathrm{e}, \mathcal{T}^\mathrm{p})$ with \eqref{eq:ew};
}
Compute $\mathfrak{E}(\Tcal^\mathrm{e},\Tcal^\mathrm{p})$ by solving  \textsf{SP-E}$(\mathcal{T}^\mathrm{e}, \mathcal{T}^\mathrm{p})$;
  \If{$\mathfrak{E}(\Tcal^\mathrm{e},\Tcal^\mathrm{p})$ \emph{is admissible}}{
Store $\mathfrak{E}(\Tcal^\mathrm{e},\Tcal^\mathrm{p})$
   }
 }
\Return{$\mathfrak{E}^\star$ \emph{with the minimum cost among all the stored $\mathfrak{E}(\Tcal^\mathrm{e},\Tcal^\mathrm{p})$}}
 \caption{Optimal relocation for general mobile storage} \label{alg:relocation}
\end{algorithm}

\subsection{Approximate optimal relocation of small mobile storage }\label{subsec:relocation-general-approx}

 The algorithm in Section~\ref{subsec:relocation-general} is conceptually simple and leverages the analytical insight derived in Section~\ref{sec:gen:mvms}. However, due to the dependence on constraint binding patterns it has a time complexity that is exponential in $T$, and therefore is not practical when the optimal relocation problem is solved over a large number of time periods. In order to eliminate this exponential dependence, we propose an alternative algorithm in the supplementary material
 which achieves a near-optimal solution and eliminates the need to know the constraint binding patterns. We discretize the state-of-charge (SoC) space to create `nodes' which have an associated SoC and location tuple. We construct a time extended graph across these nodes, and calculate edge weights based on the value of moving from one SoC and location pair to another. We then solve a longest path problem over the graph, and find that the approximate solution converges to the true optimal solution as our discretization becomes finer. The approximate algorithm can be solved in polynomial time, and the error bound is linear in the discretization step size.

\section{Illustrations}\label{sec:illustrations}



We now illustrate our results on optimal relocation for storage. In Lemma \ref{lem:arbitrage} we showed that the optimal operation of a mobile storage unit in the solution of the MPED-S problem coincides with the solution of the price arbitrage problem. We do not have access to network topology and constraints, but we do have access to historical and forecast LMPs, and we use the solution of the price arbitrage problem to illustrate the optimal operation and movement of mobile storage.

\begin{figure}
    \centering
    \includegraphics[width = 0.3\textwidth]{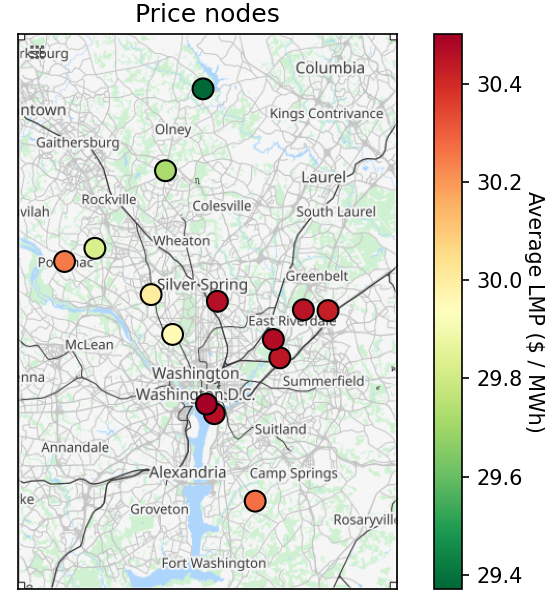}
    \caption{Average LMP of the selected nodes in May 2021}
    \label{fig:price_map}
\end{figure}

\emph{Data:} We consider a subset of price nodes in Maryland within PJM territory, shown in Fig. \ref{fig:price_map}, which are intended to be representative of LMP zones. The LMPs for nodes are obtained from the PJM DataMiner \cite{dataminer2} for the month of May 2021. LMP node names are used to identify locations on a map, and we model the entirety of an LMP zone as a single interconnection point.
The travel time between nodes is modeled using the straight line distance and a speed of 50 miles per hour, and travel cost is taken to be 4 cents/mile from the US DoE estimate \cite{EVcentspermile} for passenger EVs, and 16 cents/mile for an electric truck. 
While we do not explicitly model driver wages as a part of the cost, we will discuss the question of how to incentivize drivers to participate in Section \ref{sec:illustration-fast-charging}.
We consider a time horizon of $24$ hours, with a time step of $1$ hour.
We model two mobile storage units: a Tesla Model 3 with a 50 kWh battery and a maximum power throughput of 11 kW (level 2 AC charging power), and a Tesla Semi with a 500kWh battery and maximum power throughput of 100 kW. 

\begin{figure}
    \centering
    \includegraphics[width = 0.5\textwidth]{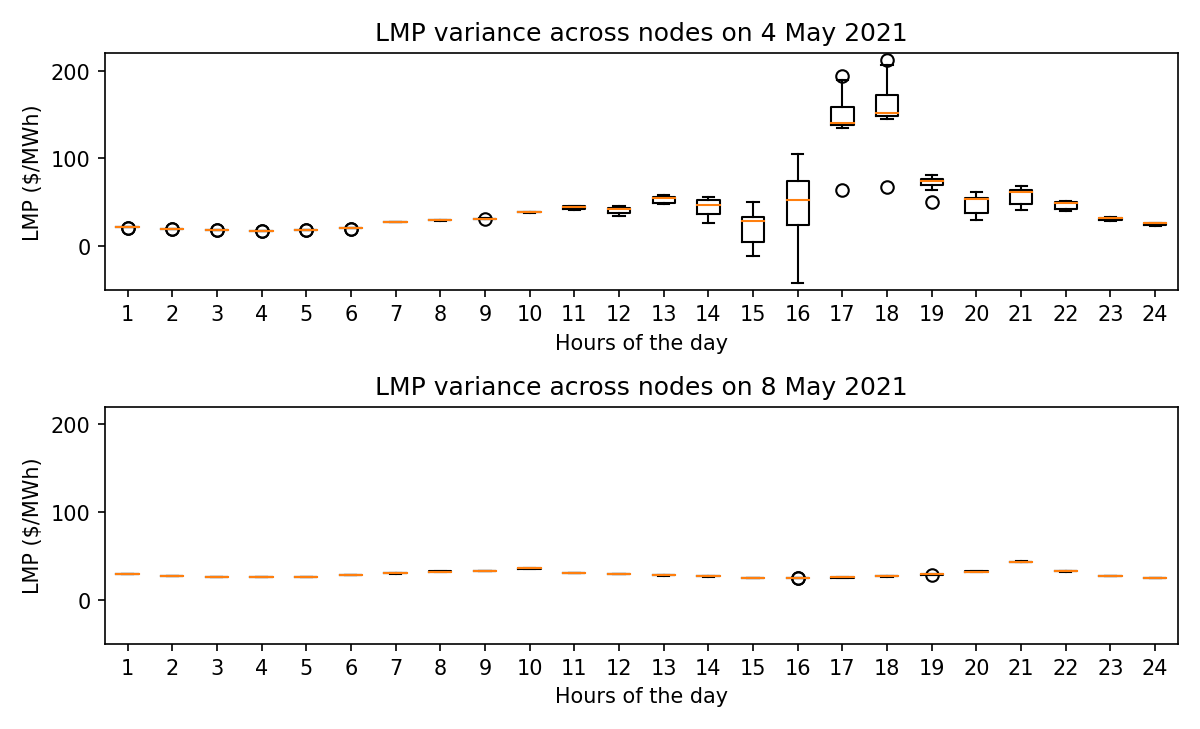}
    \caption{Range of LMPs across nodes on two dates}
    \label{fig:lmp_boxplot}
\end{figure}

The selected nodes exhibit spatial price differences, illustrated by variation in the average LMP in Fig. \ref{fig:price_map}. They also exhibit temporal price variation, i.e., the LMP at a single node varies over the course of the day. In Fig. \ref{fig:lmp_boxplot} we show the spread of prices over two days in the month of May. At each hour, the LMPs across all nodes are used to make a boxplot to illustrate how the LMPs are spread out. A longer boxplot height indicates that there is high spatial variance in LMP at that hour, which is the case on 4 May 2021 at 4 pm. A shorter boxplot indicates that the LMP across nodes is largely similar, which is the case on 8 May 2021.

The spatial variation in prices can change significantly from one day to the next, depending on the grid constraints in force on either day. A mobile storage unit, e.g., an EV can capitalize on these differences as discussed previously. We use the algorithm proposed in Section \ref{subsec:relocation-general-approx} to optimize the movement and operation of an EV over the price nodes in Fig. \ref{fig:price_map} on 4 May 2021, a day of high spatial LMP variation. 

\subsection{Tesla Model 3}

\begin{figure}
    \centering
    \includegraphics[width = 0.3\textwidth]{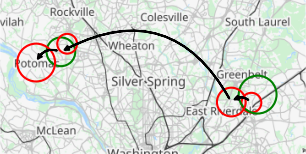}
    \caption{Movement of an EV over the course of 4 May 2021 with charging/discharging operations}
    \label{fig:movement}
\end{figure}

The EV is initially charged to $40\%$, and starts the day at the $\mathsf{LANHAM}$ node. Fig. \ref{fig:movement} illustrates the movement of the EV over the course of 4 May 2021. The arrows indicate the optimized movement of the EV over time, and the circles indicate the change in state-of-charge (SoC) at each node with a radius proportional to the magnitude of the net energy charge or discharge at that node. Red circles indicate discharging and green circles indicate charging. The EV starts at $40\%$ charge from a node in the bottom-right corner of the figure, and continues to move across nodes over the course of the day, ending the day at a node in the top-left corner with 0\% charge. 

\begin{figure}
    \centering
    \includegraphics[width = 0.5\textwidth]{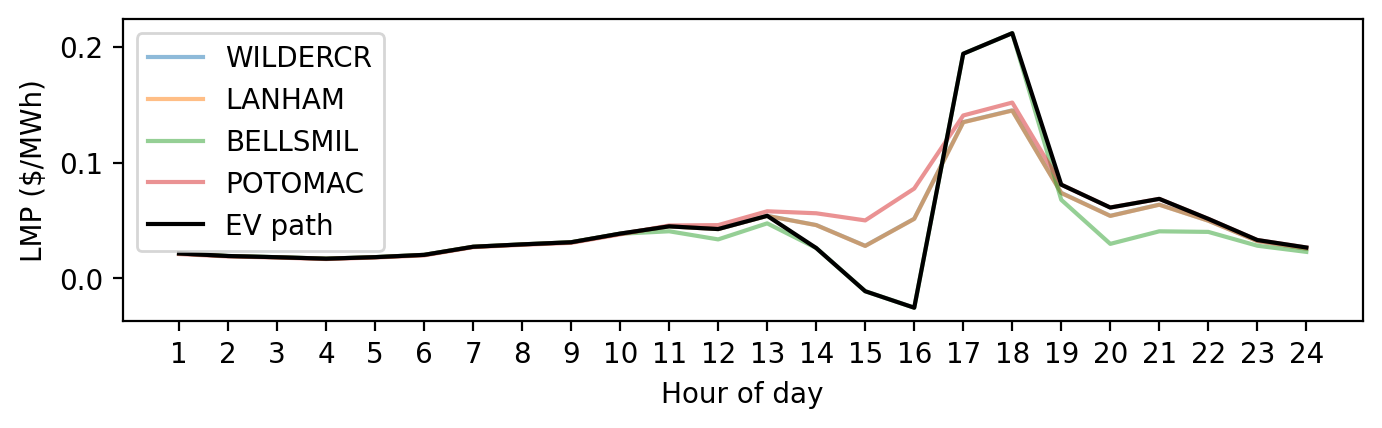}
    \caption{Effective LMPs for an EV moving along the optimal path on 4 May 2021}
    \label{fig:EVPathLMP}
\end{figure}

The EV's movement is optimized to maximize the value of mobile storage, and as it moves across nodes it experiences an effective LMP as shown in Fig. \ref{fig:EVPathLMP}. 
The value gained from arbitrage is $\$7.12$, and the travel costs work out to $\$0.78$, making the net profit $\$6.34$ for 4 May 2021. 
The EV battery is power constrained through most hours of operation, which indicates that faster bidirectional charging could enable higher value. Most of the value is captured by discharging at \textsf{BELLSMIL} node at $5$ pm and $6$ pm. This indicates that there are a few high-value hours when we can prioritize dispatch.

\begin{table}
\centering
\resizebox{\columnwidth}{!}{\begin{tabular}{|c|c|c|c|c|}
 \hline
 Node & Interval & SoC change & Value & Travel \\
 \hline
 \textsf{LANHAM} & $12$ am - $9$ am  & $30$ kWh & $-0.52 \$$ & $0.07 \$$ \\
 \textsf{WILDERCR} & $9$ am - $1$ pm  & $-30$ kWh & $1.37 \$$ & $0.62 \$$ \\
 \textsf{BELLSMIL} & $1$ pm - $6$ pm & $10$ kWh & $4.16 \$$ & $0.09 \$$ \\
 \textsf{POTOMAC} & $6$ pm - $12$ am & $-30$ kWh & $2.11 \$$ & $0 \$$ \\
 \hline
\end{tabular}}
  \caption{\label{tab:arbitrage-value} Arbitrage value and travel costs at each node.}
\end{table}

If we compare this value to the value of a stationary battery at any of the nodes, we find that though the mobile storage does better than a stationary storage located at the same starting node (\textsf{LANHAM}), but worse than a stationary battery located at \textsf{BELLSMIL}. However, a mobile storage starting at \textsf{BELLSMIL} does better than a stationary battery at the same node. The value of mobile storage comes from its mobility, but is also affected by the starting location and state-of-charge.

\begin{figure}
\begin{tikzpicture}
\begin{axis}[
    symbolic x coords={Stationary storage at \textsf{LANHAM},Stationary storage at \textsf{WILDERCR}, Stationary storage at \textsf{BELLSMIL}, Stationary storage at \textsf{POTOMAC}, Mobile storage starting at \textsf{LANHAM}, Mobile storage starting at \textsf{BELLSMIL}},
    xtick=data,
    x tick label style={font=\tiny,text width=1.2cm,align=center},
    ticklabel style = {font=\tiny},
    height = 3cm,
    width = 8.5cm,
    ylabel = {Value (\$/day)},
    label style={font=\tiny},
    y label style={at={(axis description cs:0.1,.5)},anchor=south},
    ytick distance = 0.5]
    \addplot[ybar,fill=blue] coordinates {
        (Stationary storage at \textsf{LANHAM}, 4.47)
        (Stationary storage at \textsf{WILDERCR}, 4.48)
        (Stationary storage at \textsf{BELLSMIL}, 6.43)
        (Stationary storage at \textsf{POTOMAC}, 4.77)
        (Mobile storage starting at \textsf{LANHAM}, 6.34)
        (Mobile storage starting at \textsf{BELLSMIL}, 6.978) 
    };
\end{axis}
\end{tikzpicture}
\caption{Value generated by stationary batteries compared to mobile storage starting at two different nodes.}
\end{figure}
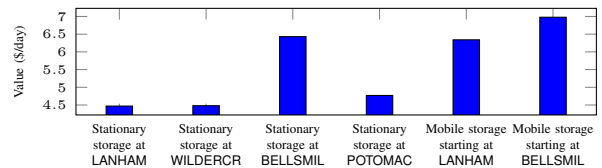


\subsection{Tesla Semi truck}

\begin{figure}
    \centering
    \includegraphics[width = 0.3\textwidth]{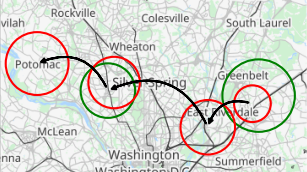}
    \caption{Movement of a Tesla Semi truck over the course of 4 May 2021 with charging/discharging operations.}
    \label{fig:semi-movement}
\end{figure}

We now consider a Tesla Semi that moves across the power network to maximize its price arbitrage value. A Tesla Semi has a significantly higher battery capacity, and can consequently generate more value. The optimal movement for a truck starting at $40\%$ capacity is illustrated in Fig. \ref{fig:semi-movement}, and the value gained from arbitrage is $69.8\$$. The travel costs work out to $3.23\$$ over the course of the day.
The net profit from operating this EV on 4 May 2021 is $66.56\$$, and the battery is power constrained for most hours through the day. This indicates that faster bidirectional charging could unlock more value from the same battery resource. Most of the value is generated by discharging during 5-7 pm, i.e., there are a few high-value hours when we can prioritize dispatch. 

\subsection{Effect of fast bidirectional charging}\label{sec:illustration-fast-charging}

In our simulation, the EV batteries are power constrained for multiple time periods and faster bidirectional charging can help generate more value. We now consider a situation where the power rating of the chargers is significantly increased. For the Tesla Model 3, we model fast charging with a 100 kW bidirectional charging speed (which is an average fast charger power throughput). For the Tesla Semi, we model fast charging with a 750 kW charge/discharge rate. These modifications ensure that the battery is almost never power constrained, and boost the net profit from arbitrage to $17.32\$$ for the Tesla Model 3 and $182.38\$$ for the Tesla Semi, starting from the same SoC and initial location as in the previous simulations. 

While putting these numerical results in perspective, we need to consider that all days will not look the same for an EV conducting price arbitrage. We chose a date which had high inter-node price variability which led to higher revenue from spatial price arbitrage. We also modeled an EV that could be controlled completely, i.e., could be moved around and charged/discharged at will. In reality, operators will likely not be able to control EV movement without giving monetary incentives, and may have to work with existing EV movement patterns (e.g., people driving their EVs to work and back). They will also not be able to charge/discharge the full battery capacity, and will have to reserve a portion of the battery to move the EV itself. Charging and discharging the battery will accelerate its degradation, and there will be associated costs that the operator will have to incur.

At the same time, there are a number of ways that mobile storage could provide value to the grid while also garnering enough compensation to incentivize the driver. As seen in Table \ref{tab:arbitrage-value}, most of the day's value is captured in a few hours, and mobile storage operators could be strategically incentivized to capture arbitrage value in those few hours. There are also a few high value nodes within the network which have a larger difference in LMPs across the day as compared to other nodes. If mobile storage is initially located near those nodes, it will be able to generate more value.
The system operator can also capitalize on existing movement patterns, e.g., people commuting from their homes to workplaces and optimize storage operation with fixed movement patterns. Larger mobile storage like battery-mounted trucks will tip the profit vs.\ cost scale further, as these resources will transport larger amounts of energy with lower labor input. A large scale simulation of EV movement patterns and LMPs across a longer timescale will help answer questions on the practicality of this scheme. 


\section{Conclusion}\label{sec:conclusion}

This paper formalizes the marginal value of mobile storage from a system operator's perspective and develops analytical expressions for two storage models: a simplified rapid storage model and a general storage model that incorporates travel time and power constraints. We developed illustrative examples to demonstrate the value of mobile storage as compared to stationary storage and wires. Efficient algorithms for the optimal storage relocation problem are then proposed based on analytical expressions for the marginal value of mobile storage. We also propose an approximate optimal relocation algorithm which is more tractable than the exact solution algorithm for the general mobile storage model. We illustrate this algorithm with a simulation of an EV moving across nodes in PJM territory, and calculate the value of optimizing operations across the power network. 

\bibliographystyle{IEEEtran}
\bibliography{bib}

\clearpage

\appendices


\section{Numerical Examples for Section~\ref{sec:econ}} \label{app:ex2data}
\paragraph*{Numerical case for Example 2}
Consider a two-period economic dispatch problem for a three bus network, with nodes $1, 2, 3$ as in Fig. \ref{f2}. Assume the system is equipped with a stationary battery of capacity $0.5$  at node $1$, and a mobile battery of capacity $0.5$  which is located at node $3$ in the first period and at node $1$ in the second period. Further, assume each node has power generation with a quadratic cost function $C_i(g_i(t)) = g_i(t)^2$, and each of the  lines has an identical susceptance  value and capacity of $0.5$. The energy demand of this system is concentrated at node $1$, with a demand of $5$  in the first period, and $10$  in the second period. In this situation, both line $2\rightarrow 1$ and line $3\rightarrow 1$ are congested in time period 1.
The LMPs for the two periods are $\lambda_1(1) = 9, \lambda_3(1) = 2, \lambda_1(2) = 16$, which means that $\textsf{MV}^\mathrm{ms}(1) = \lambda_1(2) - \lambda_3(1) = 14$, $\textsf{MV}^\mathrm{ss}(1) = \lambda_1(2) - \lambda_1(1) = 7$. Meanwhile, $\textsf{MV}^\mathrm{w}(1) = \beta_{3\rightarrow 1}(1) = 6$, and $\textsf{MV}^\mathrm{w}(1) + \textsf{MV}^\mathrm{ss}(1) = 13 < \textsf{MV}^\mathrm{ms}(1)$.

%
\paragraph*{Numerical case for Example 3}
Consider the same setup as in a) but with a  different  demand. The demand of the system in the first time period is $5$ units each at node $1$ and $2$, and $10$ units at node $1$  in the second period. In this situation, both line $3\rightarrow 1$ and line $3\rightarrow 2$ are congested in time period 1.
The LMPs for the two periods are $\lambda_1(1) = 10, \lambda_3(1) = 3, \lambda_1(2) = 16$, which means that $\textsf{MV}^\mathrm{ms}(1) = \lambda_1(2) - \lambda_3(1) = 13$, $\textsf{MV}^\mathrm{ss}(1) = \lambda_1(2) - \lambda_1(1) = 6$. Meanwhile, $\textsf{MV}^\mathrm{w}(1) = \beta_{3\rightarrow 1}(1) = 8$, and $\textsf{MV}^\mathrm{w}(1) + \textsf{MV}^\mathrm{ss}(1) = 14 > \textsf{MV}^\mathrm{ms}(1)$.


\section{Proofs for Section \ref{sec:fast}} \label{app:sec-fast-proof}

\paragraph*{Proof for Theorem \ref{th:mv-fast}}
From the Lagrangian of the economic dispatch problem in \eqref{eq:fastED}, we get that $ \textsf{MV}^\mathrm{ms}_k(\mathfrak{E}, \overline{\mathbf{s}})  = \ones^\top \bm{\mu}_k $. Further, from the stationarity KKT condition with respect to $u_k(t)$, we get 
\[
L_t^\top (\bm{\mu}_k - \bm{\nu}_k) = - \gamma(t) + E_k(t)^\top H^\top \bm{\beta}(t),
\]
where $L_t^\top$ is the $t^{th}$ column of $L^\top$, and $E_k(t)$ is the $k^{th}$ column of $E(t)$. At any time, only one of the pair of constraints in \eqref{eq:fastED:c3} will be binding. Hence, either $\mu_k(t) = 0$ or $\nu_k(t) = 0$. Using the definition of $\bm{\lambda}(t)$ from \eqref{eq:lmp}, we get the expression in \eqref{eq:mv-fast}.

\paragraph*{Proof for Lemma \ref{lem:radial}}

A radial power network can be represented as a tree, and there is a unique path between any two nodes. Consider two nodes $i, j$ with a path $i \rightarrow k_1 \rightarrow k_2 \rightarrow ... k_p \rightarrow j$, where $k_1, ... k_p$ are nodes in the power network. The LMP difference across $i,j$ at any time can be written as
\[\lambda_{j} - \lambda_i = (\lambda_j - \lambda_{k_p}) + ... + (\lambda_{k_2} - \lambda_{k_1})  +  (\lambda_{k_1} - \lambda_i), \]
i.e. the sum of LMP differences across nodes in the path between $i, j$. For any edge in the power network, we will have two $\beta$ values (one in each direction) out of which only one can be non-zero - the $\beta$ value corresponding to the direction of power flow, since the line can only be congested in that direction. From \cite{Taylor2015} we have that $\beta_{i,j}(t) - \beta_{j,i}(t) = \lambda_j(t) - \lambda_i(t)$, and if the power flow is $i \rightarrow j$, then 
\[ \textsf{MV}^\mathrm{w}_{i \rightarrow j}(t) = \beta_{i,j}(t) = \lambda_j(t) - \lambda_i(t) \geq 0.\]
The value of stationary storage located at $j$ at time $t$ is 
\[\textsf{MV}^\mathrm{ss}_j(t) =\Big( \lambda_{j}(t+1) - \lambda_{j}(t) \Big)_+, \]
where $\lambda_{j}(t+1) - \lambda_{j}(t) \geq 0$ since the stationary storage would only transfer energy across time if it had a positive value from doing so. The marginal value of mobile storage that moves from node $i$ at time $t$ to node $j$ at time $t+1$ can be expressed as 
\begin{align}
    \textsf{MV}^\mathrm{ms} =&   \Big( \lambda_{j}(t+1) - \lambda_{i}(t) \Big)_+  \\
 = &  \Big( \lambda_{j}(t+1) - \lambda_{j}(t) + \lambda_{j}(t) - \lambda_{k_p}(t) + \\
& ... + \lambda_{k_1}(t) - \lambda_{i}(t) \Big)_+ \\
= & \: \:  \textsf{MV}^\mathrm{ss}_j(t) + \textsf{MV}^\mathrm{w}_{k_p \rightarrow j} + ... + \textsf{MV}^\mathrm{w}_{i \rightarrow k_1}.
\end{align}


\section{Approximate optimal relocation of small mobile storage} 
\label{app:approx}

\subsubsection{Mobile storage operator perspective}
We first consider the optimal relocation problem from the perspective of a storage operator that knows the LMPs $\boldsymbol\lambda$ over the time horizon $T$ in advance, and wishes to optimize the operation and movement of the mobile storage. We will show that solving this problem is equivalent to solving MPED-S. At each time $t \in [T]$, say the storage unit $k$ is moved from node $i_k(t)$ to $i_k(t+1)$ and charged by $u_k(t)$.
The total value of the mobile storage over $T$ is 
\begin{align}\label{eq:operator-value}
\sum_{t = 1}^T - \lambda_{i_k(t)}(t) u_k(t) - \kappa D_{i_k(t) i_k(t+1)} ,
\end{align}
which is price arbitrage revenue net of travel costs. 
\begin{lemma}\label{lem:approx-equivalent-problem}
Suppose that for each storage $k$, the price arbitrage profit maximization problem given by 
\begin{equation}\label{eq:arbitrage-travelcost}
	\begin{aligned}
	\max_{\mathbf{u}_k} \quad &\sum_{t=1}^T - \lambda_{i_k(t)}(t) u_k(t) - \kappa D_{i_k(t) i_k(t+1)}\\
		\mbox{s.t.}\quad  &   \mathbf{0} \le L \mathbf{u}_k \le \overline{s}_k \mathbf{1},\\
		&  -  \overline{u}_k(\overline{s}_k)\bm{\Delta}^{\mathrm{S}}_k  \le \ub_k \le \overline{u}_k(\overline{s}_k)\bm{\Delta}^{\mathrm{S}}_k
	\end{aligned}
\end{equation}
for a specific path $\mathbf{i}_k$ has a unique solution. Then the optimal operation  $\mathbf{u}_k$ for storage $k$ moving along path $\mathbf{i}_k$ (defined in the $k^{th}$ column of matrices $E(t)$) in a solution of the MPED-S problem \eqref{eq:ED} is the same as the solution of the price arbitrage problem \eqref{eq:arbitrage-travelcost}.
\end{lemma}

\begin{IEEEproof}
From Lemma \ref{lem:arbitrage}, we know that the optimal solution of \eqref{eq:arbitrage} coincides with the optimal solution of the MPED-S problem \eqref{eq:ED}. \eqref{eq:arbitrage-travelcost} differs from \eqref{eq:arbitrage} in having an additional term corresponding to travel cost. However, this term is independent of the optimization variable $\mathbf{u}_k$, and does not affect the optimal solution. 
\end{IEEEproof}
The combined optimization has a mix of continuous ($\mathbf{u}_k$) and discrete ($\mathbf{i}_k$) variables, which makes it hard to solve. We now develop an algorithm that uses a discretization of the state of charge $s_k(t)$ to optimize $\ub_k, \ib_k$ in polynomial time where $u_k(t) = s_k(t) - s_k(t-1)$. 
 
 \subsubsection{Algorithm introduction and discretization set up} 
A discretization parameter $h$ is used to discretize the SoC space $s_k(t) \in [0, \bar{s}]$, such that $\hat{s}_k(t)$ is an integral multiple of $h$, with $\bar{s} = zh$ and $z \in \mathbb{N}$. The allowable states-of-charge (SoC) are now $\hat{s}_k(t) \in \{0, h, ..., zh\}$. At each time $t$, the $k^{th}$ mobile storage is located at a node $i_k(t)$ and has an SoC $\hat{s}_k(t)$.

We can couple $i_k(t), \hat{s}_k(t)$ to create $[(z+1)K]$ nodes, each of which has an SoC and location tuple associated with it. At any time, the storage is located at one of these $[(z+1)K]$ nodes. We use the $h$-floor of the continuous variable $s$ to map it to its discretized version, i.e.,
\[
\floor*{s} = mh, \quad \text{such that} \quad  mh \leq s < (m+1)h.
\] 
The algorithm is able to achieve a value that converges to the true optimum as the discretization density is increased, i.e., as the discretization parameter $h$ is made smaller. 
\subsubsection{Directed graph description} 
\begin{figure}[h]
\centering
\includegraphics[width=.45\textwidth]{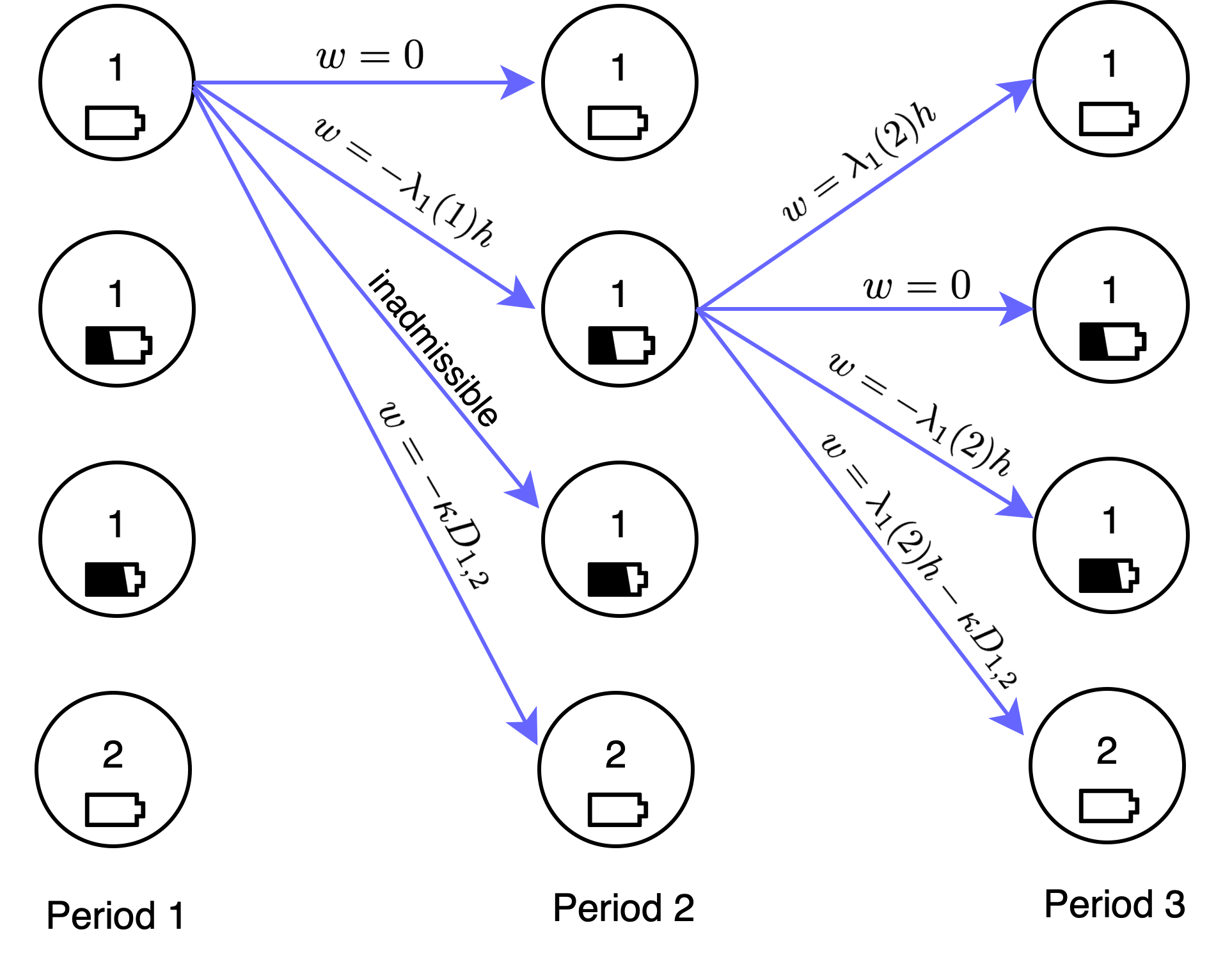}
\caption{Example of the time extended graph on which the longest path problem is defined. A sample of weights is defined on the network to illustrate the algorithm setup.}\label{fig:general-storage-network}	
\end{figure}

The value maximization problem can be solved by solving a longest-path problem over $T$. We construct a time-extended graph $G(\mathcal{V}, \mathcal{E})$ where  the set of nodes $\mathcal{V}$ includes $T$ copies of all the $[(z+1)K]$ nodes corresponding to tuples of power network locations and SoC levels, and $z+1$ dummy sink nodes, each corresponding to an SoC level in $\{0, h, ..., zh\}$. The set of edges $\mathcal{E}$ includes directed edges from every node $(i, \hat{s}_1)$ in the copy corresponding to $t$ to every node $(j, \hat{s}_2)$ in the copy corresponding to $t+1$ which satisfies \eqref{eq:ED-energy}, \eqref{eq:ED-power} with $u(t) = \hat{s}_2 - \hat{s}_1$, with an edge weight 
\begin{align}
 - \lambda_{i}(t) (\hat{s}_2 - \hat{s}_1) - \kappa D_{ij} ;
\end{align}
and directed edges from every node $(i, \hat{s}(T))$ in the $T$-th copy to dummy sink nodes $(j, \hat{s})$ which satisfy \eqref{eq:ED-energy}, \eqref{eq:ED-power} with $u(t) = \hat{s} - \hat{s}(T)$, with an edge weight 
\[
- \lambda_{i}(t) (\hat{s} - \hat{s}(T)).
\]
An example graph is illustrated in Fig. \ref{fig:general-storage-network}, and a few example weights are labeled in the figure. 

\subsubsection{Algorithm} In order to find the optimal operation and movement schedule, we solve the longest path problem 
over the directed graph. Starting from $t = T$, we assign each node a value equal to the maximum value that can be attained when starting from that node, which is the longest path (sum of edge weights). This problem can be solved in polynomial time ($O(K^2z^2T)$), but is an approximation of the actual optimal solution since it approximates a continuous variable with discrete steps. However, the difference between the discretized-state polynomial time solution and the true optimal solution can be reduced to an arbitrarily small number by reducing the discretization parameter $h$, i.e., increasing $z$. 



\subsubsection{Convergence to true optimal solution}
We will now state and prove the convergence of our discrete-space solution to the true optimal solution. 
We use the following result to prove that the discretized algorithm will converge to the true optimum.

\begin{proposition}\label{prop:floor-constraint}
Consider two optimization problems $P1$, $P2$
\begin{align}
    P1: J^{(1)} =& \max_s f(\floor*{s}) \: \: \: \: \: \text{and}  & P2: J^{(2)} = & \max_s  f(\floor*{s}) \\
    \text{s.t.} \: \:  & 0 \leq s \leq \bar{s} & \text{s.t.} \: \: & 0 \leq \floor*{s} \leq \bar{s}
\end{align}
then $J^{(1)} = J^{(2)}$.
\end{proposition}
\begin{IEEEproof}
The constraint set of $P2$ is equivalent to $0 \leq s <\bar{s} + h $, which we get by enlarging the constraint set of $P1$. Let $s_2^*$ be the optimal argument for $P2$. Then, either 
\begin{enumerate}
    \item $0 \leq s_2^* \leq \bar{s}$. In this case, $s_2^*$ satisfies the constraints of $P1$ and will be the optimal argument for $P1$ as well, and $J^{(1)} = J^{(2)}$.
    \item $\bar{s} < s_2^* < \bar{s}+h$. In this case, $J^{(2)} = f(\floor*{s_2^*}) = f(\bar{s}) \geq f(\floor*{s})$, for all $ 0 \leq s \leq \bar{s}$. Thus, the optimal argument $s_1^*$ of $P1$ will be such that $J^{(1)} = f(\floor*{s_1^*}) = f(\bar{s}) = J^{(2)}$.
\end{enumerate}
\end{IEEEproof}

To compare the discrete-space solution with the true optimal solution, we adopt the following notation. Let $J_t := J(t, s(t), i)$ be the true maximum value that can be achieved when starting with a mobile battery at time $t$ with SoC $s(t)$ and at location $i$, i.e., the true optimal solution with a continuous charge/discharge space. Here, $J_t$ is shorthand notation adopted for convenience, and $J_t$ has an underlying dependence on $s(t)$ and $i$.
Similarly, let $\hat{J_t} := \hat{J}(t, s(t), i)$ be the maximum value that can be achieved when the battery operator is only able to make discrete charge/discharge decisions, i.e., choose $\hat{u}(t) \in \{0, h, ..., zh \}$ that also satisfies operational constraints \eqref{eq:ED-energy}, \eqref{eq:ED-power}, where $\floor*{s(t)}$ is the discrete SoC. We now analyze the difference between the two solutions at time $t=1$, i.e., at the start of the decision horizon. 

\begin{theorem}\label{thm:approximate-convergence}
Let $J_1$ denote the maximum value that can be achieved by the mobile storage starting with state of charge $s(1)$ at node $i(1)$, and let $\hat{J_1}$ denote the maximum value that can be achieved with the same initial conditions but discrete charge/discharge decisions. Then 
    \begin{align}\label{eq:upper-bound}
        |J_1 - \hat{J_1}| \leq  
        h \left[ |\lambda_{i(1)}(1)| + \sum_{\tau = 2}^T \max_{i(\tau)} |\lambda_{i(\tau)}(\tau)|  \right],
    \end{align}
    for each possible $s(1)$ and $i(1)$, and the right hand side converges to $0$ as the discretization step size $h \rightarrow 0$.
    \end{theorem}

\begin{IEEEproof}
 At each time step, the battery operator can move the battery to a new location $j$, and charge the battery by any amount $u(t)$ that satisfies the battery operational constraints \eqref{eq:ED-energy}, \eqref{eq:ED-power}.
By definition, 
\begin{equation}
    |s(t) - \floor*{s(t)}| \leq h.
\end{equation}
We prove the statement in \eqref{eq:upper-bound} using induction. First, note that 
\begin{align}
    &J(T,s(T), i(T)) = - \lambda_{i(T)}(T) \max \{ -s(t), -\bar{u}(\bar{s})\}, \; \text{and} \\
    &\hat{J}(T,s(T), i(T)) = - \lambda_{i(T)}(T) \max \{ - \floor*{s(t)}, -\bar{u}(\bar{s})\}, 
\end{align}
as the battery operator will choose to maximize the value by discharging at the maximum possible rate at a positive LMP, as it can derive no value from the remaining charge after $T$. Then we have that
\begin{align}
    |J_T - \hat{J}_T | \leq |\lambda_{i(T)}(T)|h.
\end{align}
For time $T-1$, consider a state of charge $s(T-1)$. The maximum value that can be attained through continuous battery operation is given by 
\begin{align}
     & J(T-1, s(T-1), i(T-1)) \\
     & \quad \quad \quad = \;    \max_{i(T)}   \max_{s(T)}  J(T, s(T), i(T)) - \kappa D_{i(T) i(T-1)} \\
      & \quad \quad \quad  \quad \quad \quad \quad -  \lambda_{i(T-1)}(T-1) \{s(T) - s(T-1)\} \\
       & \quad \quad \quad \quad \quad \text{s.t.} \: u = s(T) - s(T-1) \: \text{satisfies} \: \eqref{eq:ED-energy}, \eqref{eq:ED-power} , 
\end{align}
and the maximum value attainable through discrete battery operation is given by
\begin{align}
    & \hat{J}(T-1, s(T-1), i(T-1))  \\
    & \quad \quad \quad = \; \max_{i(T)}  \max_{s(T)} \hat{J}(T, s(T), i(T)) - \kappa D_{i(T) i(T-1)} \\
     & \quad \quad \quad  \quad \quad \quad \quad  -  \lambda_{i(T-1)}(T-1) \{\floor*{s(T)} - s(T-1)\} \\
     & \quad \quad \quad  \quad \quad \text{s.t.} \: \: \hat{u} = \floor*{s(T)} - s(T-1) \; \text{satisfies} \; \eqref{eq:ED-energy}, \eqref{eq:ED-power}.
\end{align}
Then define the optimal solutions to the inner optimization problems as 
\begin{align}
   & g(T-1, s(T-1), i(T-1))  \\
   & \quad =  \max_{s(T)} J(T, s(T), i(T))  -  \lambda_{i(T-1)}(T-1) s(T)  \\
     & \quad \quad \quad \text{s.t.} \: \: u = s(T) - s(T-1) \; \text{satisfies} \; \eqref{eq:ED-energy}, \eqref{eq:ED-power}, \; \text{and} \\
    & \hat{g}(T-1, s(T-1), i(T-1)) \\
    & \quad =  \max_{s(T)} \hat{J}_T -  \lambda_{i(T-1)}(T-1)\floor*{s(T)}  \\
    & \quad \quad \quad  \text{s.t.} \: \: \hat{u} = \floor*{s(T)} - s(T-1) \; \text{satisfies} \; \eqref{eq:ED-energy}, \eqref{eq:ED-power} \\
    & \quad =  \max_{s(T)} \hat{J}_T -  \lambda_{i(T-1)}(T-1)\floor*{s(T)}  \\
     & \quad \quad \quad \text{s.t.} \: \: u = s(T) - s(T-1) \; \text{satisfies} \; \eqref{eq:ED-energy}, \eqref{eq:ED-power}
\end{align}
from Proposition \ref{prop:floor-constraint}. Then
\begin{align}
     & |\hat{g}_{T-1} - g_{T-1}|   \\
   &  \leq  \max_{s(T)} |  J_T - \hat{J}_T   -\lambda_{i(T-1)}(T-1) (s(T) - \floor*{s(T)})| \\
    &  \quad \quad \text{s.t.} \: \: u = s(T) - s(T-1) \; \text{satisfies} \; \eqref{eq:ED-energy}, \eqref{eq:ED-power} \\
   &  \leq   |\lambda_{i(T-1)}(T-1)|h + |\lambda_{i(T)}(T)|h
\end{align}
using the fact that for two functions $f(s)$ and $ g(s)$, 
\begin{align}
    | \max_s f(s) - \max_s g(s) | \leq \max_s |f(s) - g(s)|, 
\end{align}
and the triangle inequality. This leads to 
\begin{align}
    &|J_{T-1} - \hat{J}_{T-1}| \\
    & = \left| 
    \max\limits_{i(T)} \left( \begin{array}{l}g_{T-1}  \\
    - \kappa D_{i(T) i(T-1)} 
    \end{array} \right) -  \max\limits_{i(T)} \left( \begin{array}{l}  \hat{g}_{T-1}  \\
     - \kappa D_{i(T) i(T-1)}
     \end{array}\right) \right|  \\
    & \leq \max_{i(T)} |g_{T-1} - \hat{g}_{T-1}|\\
    & = |\lambda_{i(T-1)}(T-1)|h + \max_{i(T)} |\lambda_{i(T)}(T)|h
\end{align}
which satisfies our induction statement. Now assuming that the statement holds for a specific $t$, we will show that it also holds for $t-1$. We know from our induction hypothesis that 
\begin{align}
    |J_{t} - \hat{J}_{t}| \leq  h \left[ |\lambda_{i(t)}(t)| + \sum_{\tau = t+1}^T \max_{i(\tau)} |\lambda_{i(\tau)}(\tau)|  \right]
\end{align}
Defining $J_{t-1
}, \hat{J}_{t-1}, g_{t-1}, \hat{g}_{t-1}$ like their analogues for $T-1$, we have 
\begin{align}
    |\hat{g}_{t-1} - g_{t-1}| \leq & h  |\lambda_{i(t-1)}(t-1)| \\
   & + h \left[  |\lambda_{i(t)}(t)| + \sum_{\tau = t+1}^T \max_{i(\tau)} |\lambda_{i(\tau)}(\tau)|  \right]
\end{align}
which gives us
\begin{align}
    |J_{t-1} - \hat{J}_{t-1}|& \leq \max_{i(t)} |g_{t-1} - \hat{g}_{t-1}| \\
    & = \left[|\lambda_{i(t-1)}(t-1)| h +  \sum_{\tau = t}^T \max_{i(\tau)} |\lambda_{i(\tau)}(\tau)| \right]  
\end{align}
thus proving our induction hypothesis.

\end{IEEEproof}

\section{Proofs for Section \ref{sec:general}} \label{app:sec-general-proof}

\subsubsection{Proof for Lemma \ref{lem:arbitrage}}
Consider the Lagrangian of the MPED-S problem, specifically the terms involving $\ub_k$
\begin{align}
    \mathcal{L} = f(\mathbf{p}) + \sum_k \min_{\ub_k} & \sum_t \lambda_{i_k(t)}(t) u_k(t)  + \boldsymbol\mu_k^\top (L \ub_k - \overline{s}_k)  \\
    & - \boldsymbol\nu_k^\top L \ub_k + \boldsymbol\omega_k^\top (-\ub_k - \overline{u}_k(\overline{s}_k)\bm{\Delta}^{\mathrm{S}}_k ) \\
    & + \boldsymbol\phi_k^\top (\ub_k - \overline{u}_k(\overline{s}_k)\bm{\Delta}^{\mathrm{S}}_k ) .
\end{align}
The Lagrangian can be decomposed over $\ub_k$, and each inner minimization problem over $\ub_k$ is equivalent to the Lagrangian of \eqref{eq:arbitrage}, which proves that their optimal solutions will coincide.

\subsubsection{Proof for Theorem \ref{th:mv-general}}
From the Lagrangian of the economic dispatch problem in \eqref{eq:ED}, we get that $\textsf{MV}^\mathrm{ms}_k(\mathfrak{E}, \mathbf{\bar{s}}) =  \mathbf{1}^\top \bm{\mu}_k + \bar{u}_k'(\bar{s}_k) (\bm{\Delta}^{\mathrm{S} }_{k})^\top (\bm{\omega}_k + \bm{\phi}_k)$. Further, from the stationarity KKT condition with respect to $u_k(t)$, we get 
\[
L^\top (\bm{\mu}_k - \bm{\nu}_k) + \bm{\omega}_k - \bm{\phi}_k = - \bm{\lambda}_{i_k}.
\]
Under Assumption \ref{ass:licq}, only one of the dual variables is non-zero at any given time. We can construct a dummy variable $\bm{z}_k = \bm{\mu}_k - \bm{\nu}_k + \bm{\omega}_k - \bm{\phi}_k$ and $T \times T$ dimensional diagonal matrices $A, B$ such that $A\bm{z}_k = \bm{\mu}_k - \bm{\nu}_k ; B \bm{z}_k  = \bm{\omega}_k - \bm{\phi}_k$. Matrix $A$ has diagonal entries equal to $1$ corresponding to times when the capacity constraint is active, and $0$ otherwise. Similarly, $B$ has diagonal entries equal to $1$ corresponding to times when the power constraint is active, and $0$ otherwise, i.e., $A+B = I$. 
As defined previously, $L$ is a lower triangular matrix defined as $L_{ij} = 1$ if $i \geq j$, and $0$ otherwise. $L^\top$ then becomes an upper triangular matrix. The matrix $L^\top A$ has a $1$ at indices $(t_1,t_2)$ where $t_1 \leq t_2$ and $A_{t_2 t_2} = 1$, i.e., if the energy capacity constraint is active at time $t = t_2$. This ensures that $L^\top A$ has a rank equal to the rank of matrix $A$, since it has non-zero linearly independent rows for all times when the energy capacity constraint is active. Since $B$ has linearly independent rows for power capacity constrained periods, we can see that $L^\top A + B$ is full rank with all the diagonal elements being $1$, and the elements at $(t_1, t_2)$ being $1$ if $t_1 \leq t_2$ and $A_{t_2 t_2} = 1$. Then, we can obtain an expression for $\bm{z}_k$
\[
\bm{z}_k = (L^\top A + B)^{-1} (-\bm{\lambda}_{i_k}).
\]
It is easy to verify that the structure of matrix $(L^\top A + B)^{-1}$ is such that the $t_p$-th column (corresponding to a power constrained time) is an elementary vector $\bm{e}_{t_p}^\top$, and the $t_e$-th column (corresponding to an energy capacity constrained time) is $[-1, ..., -1, 1, 0, ..., 0]^\top$ where $-1$ appears $t_e - 1$ times, and the $1$ is at the $t_e$-th location. 
We can use this equation to  obtain expressions for $\bm{\mu}_k - \bm{\nu}_k, \bm{\omega}_k - \bm{\phi}_k$. Replacing these expressions into the marginal value formulation leads us to the result in Theorem \ref{th:mv-general}.

\end{document}